\documentclass[10pt]{article}

\usepackage{ragged2e}       
\usepackage{soul}           
\usepackage{xcolor}
\usepackage{graphicx}
\usepackage{hyperref}
\usepackage{amsmath}
\usepackage{subcaption}
\usepackage{amssymb}
\usepackage{booktabs}
\usepackage{fullpage}
\usepackage[section]{placeins}
\usepackage{authblk}
\usepackage[round]{natbib}

\definecolor{my_purple}{RGB}{255,0,255}
\graphicspath{{Figures_Peristaltic_Regimes/}}

\definecolor{REDCOLOR2}{RGB}{0,0,0}

\title{\textbf{\Large Peristaltic regimes in esophageal transport}}

\author[1]{\normalsize Guy Elisha}
\author[1]{\normalsize Shashank Acharya}
\author[2]{\normalsize Sourav Halder}
\author[3]{\normalsize Dustin A. Carlson}
\author[3]{\normalsize Wenjun Kou}
\author[3]{\normalsize Peter J. Kahrilas}
\author[3]{\normalsize John E. Pandolfino}
\author[1,2]{\normalsize Neelesh A. Patankar\thanks{Corresponding author: N.~A.~Patankar (\texttt{n-patankar@northwestern.edu})}}

\affil[1]{Department of Mechanical Engineering, McCormick School of Engineering, \newline
Northwestern University Technological Institute, 2145 Sheridan Road, Evanston, IL 60201 \vspace{1ex}}
\affil[2]{Theoretical and Applied Mechanics Program, McCormick School of Engineering,\newline
Northwestern University Technological Institute, 2145 Sheridan Road, Evanston, IL 60201 \vspace{1ex}}
\affil[3]{Division of Gastroenterology and Hepatology, Feinberg School of Medicine, \newline
Northwestern University, 676 North St. Clair Street, Arkes Suite 2330, Chicago, IL 60611 \vspace{1ex}}
\date{}

\begin{document}

\captionsetup[figure]{labelfont={bf},name={Fig.},labelsep=space}

\maketitle 

\begin{abstract}
A FLIP device gives cross-sectional area along the length of the esophagus and one pressure measurement, both as a function of time. Deducing mechanical properties of the esophagus including wall material properties, contraction strength, and wall relaxation from these data is a challenging inverse problem. Knowing mechanical properties can change how clinical decisions are made because of its potential for in-vivo mechanistic insights. To obtain such information, we conducted a parametric study to identify peristaltic regimes by using a 1D model of peristaltic flow through an elastic tube closed on both ends and also applied it to interpret clinical data. The results gave insightful information about the effect of tube stiffness, fluid/bolus density and contraction strength on the resulting esophagus shape through quantitive representations of the peristaltic regimes. Our analysis also revealed the mechanics of the opening of the contraction area as a function of bolus flow resistance. Lastly, we concluded that peristaltic driven flow displays three modes of peristaltic geometries, but all physiologically relevant flows fall into two peristaltic regimes characterized by a tight contraction.
\end{abstract}


\section{Introduction}
The esophagus is a tubular organ that connects between the mouth and the stomach \citep{Mittal2016}. A healthy functioning esophagus transports swallowed material towards the stomach by a peristaltic contraction wave which leads to an increase in pressure distal of contraction \citep{Regan2012, Jain2019}. Understanding this peristaltic behavior and deducing the mechanical properties of the esophagus can change how clinical decisions are made. Being able to quantify the parameters that define a healthy swallow can contribute to the process of classifying and diagnosing esophageal disease progression. Common esophageal disorders include gastroesophageal reflux disease (GERD), achalasia and eosinophilic esophagitis (EoE) \citep{Aziz2016}. One tool that is used today to determine the state of the esophagus is the functional lumen imaging probe (FLIP) \citep{Jain2019, Carlson2015}. The FLIP is composed of a flexible catheter enclosed by a bag. It is placed within the esophageal lumen and records the esophagus wall response to the filling of the FLIP bag. When the bag volume increases, it activates distention-induced contractions of the esophagus wall \citep{Regan2012}. The FLIP records the esophagus cross-sectional area at different locations and one distal pressure measurement as a function of time. However, deducing the mechanical properties from FLIP data is a complicated inverse problem.

\begin{figure*}

    \centering
    \begin{subfigure}[b]{0.7\textwidth}
        \centering
        \includegraphics[trim=75 350 60 350,clip,width=\textwidth]{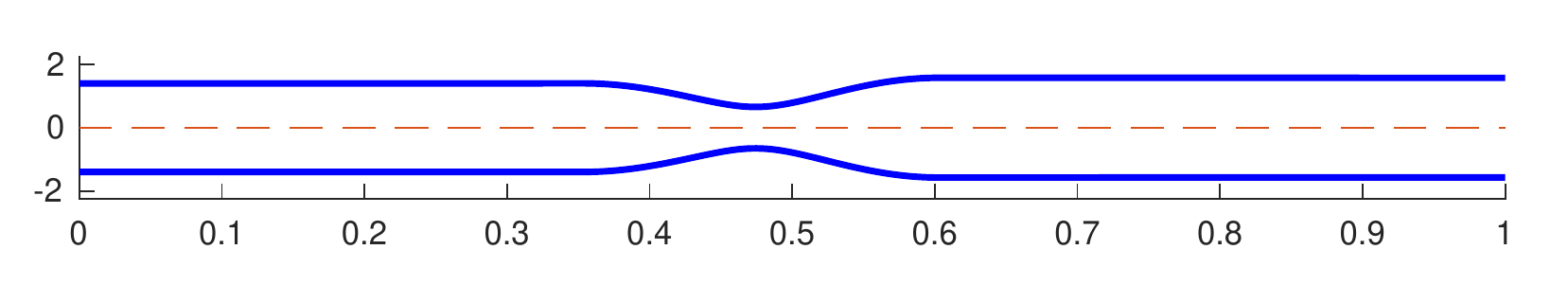}
        \caption{Regime 1}
        \label{fig:regime1_demo}
    \end{subfigure}
    \ 
    \begin{subfigure}[b]{0.7\textwidth}  
        \centering 
        \includegraphics[trim=75 350 60 350,clip,width=\textwidth]{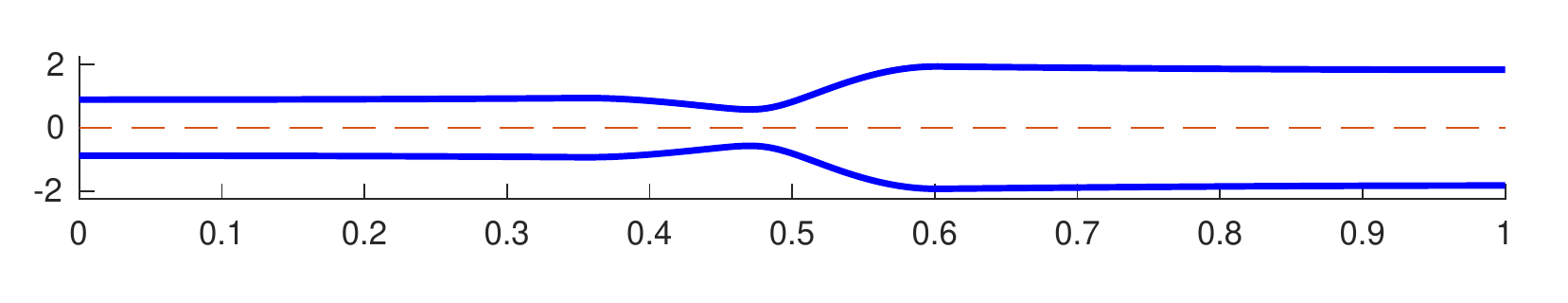}
        \caption{Regime $2$}
        \label{fig:regime2_demo}
    \end{subfigure}
    \ 
    \begin{subfigure}[b]{0.7\textwidth}   
        \centering 
        \includegraphics[trim=75 350 60 350,clip,width=\textwidth]{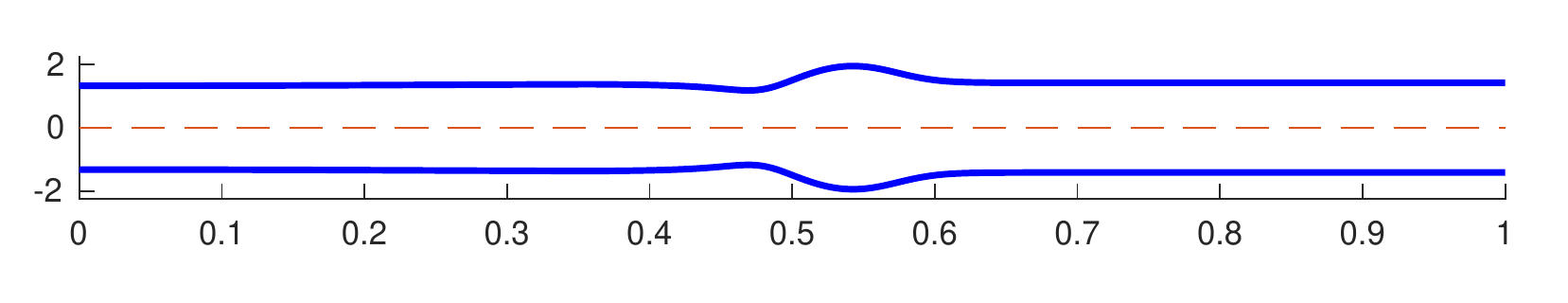}
        \caption{Regime 3}
        \label{fig:regime3_demo}
    \end{subfigure}
    \ 
    \caption{Tube geometries of the three physiologically relevant peristaltic regimes identified by \cite{Acharya_2021}. The shapes were captured from three different simulations at a single time instance in the contractile cycle.}
    \label{fig:regimes}
\end{figure*}

Extensive studies were conducted to better understand peristaltic flow through a tubular geometry \citep{Jaffrin1971, Fung1968, Burns1967, Latham1966PhDThesis}. However, limited work was done to study peristaltic flow through an elastic tube closed on both ends. In a study by \cite{Acharya_2021}, a 1D model of a peristaltic flow through an elastic closed tube was developed in order to investigate the relationship between tube stiffness, fluid viscosity, peristaltic contraction strength, internal pressure in the bag, and the resulting tube geometry. The model intended to imitate a flow inside a FLIP device located in the esophagus. Three peristaltic regimes were identified based on the resulting shape of the elastic tube which helped to better understand the different elements of peristaltic flow \citep{Acharya_2021}. The geometries of the three regimes are presented in figure~\ref{fig:regimes}. As concluded by \cite{Acharya_2021}, in the process of moving from peristaltic regime $1$ to $3$, the viscosity is increased such that it creates a competition between the force from the wall of the tube and the resistance to the flow.

The separation into regimes can be a useful tool in clinical practice as it allows us to differentiate between effective and ineffective distention-induced peristalsis. Regime $1$, presented in figure~\ref{fig:regime1_demo}, is an example of an ineffective case. As contraction goes down the deformable tube, some of the fluid flows back through the contraction instead of moving forward towards the distal end of the tube. Regime $2$, presented in figure~\ref{fig:regime2_demo}, on the other hand is an example of an effective case. In tubes that display regime $2$, flow resistance is high and the tube wall distal of contraction expands so that the peristaltic contraction wave carries the fluid forward. Lastly, peristaltic regime $3$, presented in figure~\ref{fig:regime3_demo} is another example of an ineffective case. In tubes that display regime $3$, the area downstream grows fast to an extent that makes it hard for the peristaltic contraction wave to push the fluid forward \citep{Takagi2011,Acharya_2021}, forcing the contraction to open and allow back flow. 

The peristaltic regimes discussed in \cite{Acharya_2021} were identified qualitatively, based on the geometry of the tube. In this work, we aimed to develop a metric which allows us to quantify and distinguish the different regimes numerically, and therefore, get a deeper understanding of the reasons behind their formation. Moreover, through the proposed mathematical representations of the regimes, we explained the physics behind the transition between peristaltic regimes 2 and 3 and obtained insightful information on the leading parameters that cause this transition. Lastly, we applied the proposed metric to clinical FLIP data to learn about their peristaltic behaviors and identify their peristaltic regimes. 

\section{Problem Formulation \& Numerical Solution} \label{problemSetUp}


\subsection{Governing Equations in 1D} \label{ReducedOrderDetails}

The problem considers a 1D peristaltic flow of an incompressible, viscous, newtonian fluid through an elastic tube closed on both ends. The mass conservation equation is
\begin{equation} \label{eq:continuity}
    \frac{\partial A}{\partial t}+\frac{\partial\left(Au\right)}{\partial x} = 0,
\end{equation}
and the momentum conservation equation is
\begin{equation} \label{eq:momentum}
    \frac{\partial u}{\partial t} + u\frac{\partial u}{\partial x} = 
    -\frac{1}{\rho}\frac{\partial P}{\partial x}-\frac{f_D}{\rho}.
\end{equation}
Here, $A(x,t)$ is the tube cross-sectional area, $u(x,t)$ is the fluid velocity (averaged at each cross-sectional area), $\rho$ is fluid density, $P$ is the pressure inside the tube, and $f_D$ is friction due to drag. These 1D forms of the continuity and the momentum conservation equations were derived by \cite{Ottesen2003} and have been widely used to describe valveless pumping \citep{Acharya_2021,Manopoulos2006, Bringley2008}. 

In addition, a linear, constitutive relation between pressure and the cross-sectional area of the tube is introduced, such that
\begin{equation}
\frac{\Delta{P}}{\rho}=\frac{K_{\scriptscriptstyle e}}{\rho}\left(\frac{A}{A_{\scriptscriptstyle o}}-1\right),
\label{eqn: tube_law_init}
\end{equation}
where $K_e$ is tube stiffness and $A_o$ is the undeformed reference area (or rest area) representing the cross-sectional area of the tube when ${\Delta{P}}=0$. The pressure term on the left hand side of equation (\ref{eqn: tube_law_init}) represents the difference between the pressure inside and outside the tube, such that $\Delta{P}=P_i-P_o$. Since the pressure outside of the tube is assumed to be the reference pressure (zero), $\Delta{P}=P_i=P$, we can express the pressure inside the tube in terms of area, such that
\begin{equation}
\frac{P(x,t)}{\rho}=\frac{K_{\scriptscriptstyle e}}{\rho}\left(\frac{A(x,t)}{A_{\scriptscriptstyle o}}-1\right).
\label{eqn: tube law no Act}
\end{equation}
A derivation of this linear relation is available in \citep{Whittaker2010}, and it has been verified experimentally by \cite{Kwiatek2011}. 

The esophagus transports fluid through a peristaltic contraction wave which is modeled to be sinusoidal with time. Simultaneously, the esophagus wall distal of contraction relaxes, making the wall more compliant at that region. The relaxation segment travels sinusoidally with time distal of contraction \citep{Mittal2016,AbrahaoJr2010}. Therefore, in order to mimic the contraction and relaxation of the muscle fibers of the esophagus, an activation term $\theta=\theta(x,t)$ is introduced \citep{Acharya_2021,Ottesen2003,Manopoulos2006, Bringley2008}, which is responsible for the changes in the reference area of the tube wall, such that
\begin{equation}
\frac{P(x,t)}{\rho}=\frac{K_{\scriptscriptstyle e}}{\rho}\left(\frac{A(x,t)}{A_{\scriptscriptstyle o}\theta(x,t)}-1\right).
\label{eqn: tube law}
\end{equation}
This approach is commonly used when the external activation pressure at a specific location varies sinusoidally with time. Contraction is associated with the reduction of the tube area implying $\theta<1$, and relaxation is associated with increase in the tube area implying $\theta>1$. For $\theta=1$, neither contraction nor relaxation are present. The expression for $\theta(x,t)$ and its values are elaborated upon in section \ref{Peristaltic_Wave}.

\subsubsection{Non-dimensionalizing Dynamic Equations} \label{nondimParametersGov}

To obtain a numerical solution for the peristaltic driven flow inside a closed, deformable tube, equations (\ref{eq:continuity}), (\ref{eq:momentum}) and (\ref{eqn: tube law}) were non-dimetionalized as proposed by \cite{Acharya_2021} using
\begin{equation}
A=\alpha A_{o}, \qquad t=\tau\frac{L}{c_w}, \qquad u=Uc_w, \qquad P=pK_e, \qquad \text{and} \qquad x=\chi L.
\end{equation}
Here, $c_w$ is the speed of the peristaltic wave, $L$ is the length of the tube, and $\alpha$, $\tau$, $U$, $p$, and $\chi$ are non-dimensional variables of area, time, velocity, pressure, and position, respectively. The purpose of this step is to reduce the number of independent variables. Moreover, non-dimensionalization is convenient for a parametric study of the problem. The non-dimensional form of equations (\ref{eq:continuity}) and (\ref{eq:momentum}), are

\begin{equation} \label{eq:continuity_nondim}
    \frac{\partial\alpha}{\partial\tau}+\frac{\partial\left(\alpha U\right)}{\partial\chi} = 0 , \qquad \text{and}
\end{equation}

\begin{equation} \label{eq:momentum_nondim_Gamma}
    \frac{\partial U}{\partial\tau} + U\frac{\partial U}{\partial\chi} + 
    \psi\frac{\partial p}{\partial\chi}
    + \Gamma = 0,
\end{equation}

\noindent respectively, where  $\psi=K_e/(\rho c_w^2)$  and $\Gamma =  f_D{L}/{\rho c_w^2}$ \citep{Acharya_2021}. Note that the friction term $f_D$ is a function of both the cross-sectional area of the tube and fluid velocity. Assuming a parabolic flow inside the tube, the term $f_D$ is approximated as $f_D=8\pi\mu u/(\rho A)$ . Therefore, the non-dimensional variable $\Gamma$ can be written as $\Gamma =\beta\frac{U}{\alpha}$ where $\beta = 8\pi\mu L/(\rho A_o c_w)$. Plugging this into equation \ref{eq:momentum_nondim_Gamma} we obtain

\begin{equation} \label{eq:momentum_nondim}
    \frac{\partial U}{\partial\tau} + U\frac{\partial U}{\partial\chi} + 
    \psi\frac{\partial p}{\partial\chi}
    + \beta\frac{U}{\alpha} = 0.
\end{equation}

\noindent The non-dimensional form of equation (\ref{eqn: tube law}), is

\begin{equation} \label{eq:tubelaw_nondim}
    \textcolor{REDCOLOR2}{p=\left( \frac{\alpha}{\theta}-1 \right)- \eta\frac{\partial\left(\alpha U \right)}{\partial\chi}}
\end{equation}

\noindent \citep{Acharya_2021}. Notice that this non-dimensional equation for pressure includes the term $\eta\frac{\partial\left(\alpha U \right)}{\partial\chi}$ on the right hand side, which does not appear in the dimensional form of the equation. This term is added in order to regularize the system of equations which helps stabilizing the numerical solution \citep{Acharya_2021,Wang2014}. This term is also known as the damping term, which is kept small in order to approximate the original the pressure-area equation. The damping term's dimensional form is $Y\frac{\partial \left(Au\right)}{\partial x}$, where Y is the damping coefficient. Therefore, $\eta=(Yc_wA_o)/(K_eL)$.

Lastly, we can reduce our system into two equations by substituting the non-dimensional pressure equation (\ref{eq:tubelaw_nondim}) into the non-dimensional momentum equation (\ref{eq:momentum_nondim}) to obtain

\begin{equation} \label{eq:momentum_nondim_final}
\frac{\partial U}{\partial\tau} + U\frac{\partial U}{\partial\chi} + \beta\frac{U}{\alpha} + 
        \psi\frac{\partial}{\partial\chi}\left(\frac{\alpha}{\theta}\right) = 
        \zeta\frac{\partial^2}{\partial\chi^2}\left(\alpha U\right),
\end{equation}
\noindent where $\zeta$ is equal to the product of $\eta$ and $\psi$.

\subsection{Peristaltic Wave Input and Active Relaxation} \label{Peristaltic_Wave}

As previously mentioned, $\theta=\theta(\chi,\tau)$ is the activation term introduced in the constitutive equation for pressure to resemble the contraction and relaxation of the esophagus' muscles. This term is responsible for the changes in the reference cross-sectional area of the tube wall as the wave travels in time along the length of the tube. The pattern used for $\theta$ is based on clinical study on esophageal peristalsis \citep{Goyal2008,Crist1984}. Recall that contraction is associated with the reduction of the tube reference area and relaxation is associated with increase in the tube reference area. In order to implement these processes in the model, the activation term $\theta(\chi,\tau)$ is defined by a step-wise wave function \citep{Acharya_2021,Ottesen2003,Manopoulos2006, Bringley2008} of the form

\begin{equation}\label{eq:activation_theta}
\theta(\chi,\tau)=
  \begin{cases}
  \theta_2, & \chi<\tau-w\\
  \theta_2 - \left(\frac{\theta_2-\theta_{c}}{2}\right)  \Big[1 \ +    
\sin\left(\frac{2\pi}{w}\left(\chi-\tau\right) +
        \frac{3\pi}{2}\right)\Big], & \tau-w\leq\chi\leq\tau \\
          \theta_2 - \left(\frac{\theta_2-\theta_{1}}{2}\right)  \Big[1 \ +  
\sin\left(\frac{2\pi}{w_R}\left(\chi-\tau-w_R\right) +
        \frac{3\pi}{2}\right)\Big], & \tau<\chi\leq\tau+\frac{w_R}{2} \\
  \theta_1, & \tau+\frac{w_R}{2}<\chi,
\end{cases}
\end{equation}

\noindent where $w$ is the non-dimensional width of the peristaltic wave defined by $w =
W/L$, with $W$ denoting the dimensional contraction width, and $w_R$ represents the width of the wave connecting between the contraction and relaxation. Note that the non-dimensional wave speed is 1.0 since, as in \citep{Acharya_2021}, the velocity scale was chosen to be the same as the speed of the peristaltic wave.

\begin{figure*}
    \centering{\fbox{\includegraphics[clip,width=1.0\textwidth]{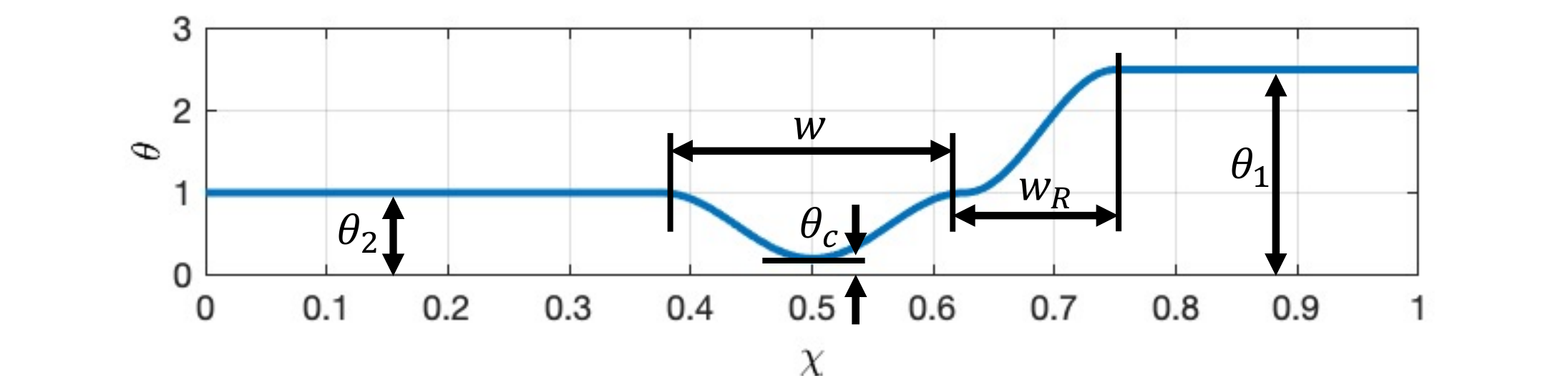}}}
    \caption{A plot of the peristaltic activation function $\theta$ from equation (\ref{eq:activation_theta}) along the tube length at a time instant. Contraction occurs when $\theta<1$ and relaxation occurs when $\theta>1$. }
    \label{fig:activation_function}
\end{figure*}

Figure \ref{fig:activation_function} presents a plot of the activation function $\theta(\chi,\tau)$ at a time instant, which helps to visualize equation (\ref{eq:activation_theta}). As concluded by \cite{AbrahaoJr2010}, active relaxation is observed only distal of contraction. Therefore, proximal of contraction (location $2$), $\theta_2=1$, implying that there is no active contraction or relaxation at that location. Distal of contraction (location $1$), $\theta_1\geq1$ implying that there is active relaxation. The values of $\theta_1$ examined in this work are $1.0$, $2.5$, and $5.0$, which were chosen based on the results obtained by \cite{Halder_2021}. At location $c$, $\theta$ is the smallest, which represents the point on the tube where the contraction is strongest. The value of $\theta$ at location $c$ is noted at $\theta_c$, and is called the contraction strength. The smaller the value of $\theta_c$, the tighter the contraction. The physiologically relevant values of $\theta_c$ that are considered in this work have been proposed by \cite{Acharya_2021}, and are $0.05$, $0.10$, $0.15$, and $0.20$. Note that the smaller the value of $\theta_c$, the tighter the contraction and the better the peristaltic wave is in pushing fluid forward. 

\subsection{Boundary Conditions} \label{Boundary Conditions}

After identifying and non-dimensionalizing the system of equations that define the problem of interest, and stating the activation forces that drive the flow, one needs to formulate boundary conditions in order to solve for the non-dimensional area ($\alpha$) and velocity ($U$). When $\eta$ in equation (\ref{eq:tubelaw_nondim}) is equal to zero, $\zeta$ is equal to zero, and the system of equations is hyperbolic \citep{Acharya_2021}. This allows only one boundary condition for $\alpha$ and one for $U$. However, recall that the tube is closed on both ends, and the volume inside the tube remains constant. Therefore, the boundary conditions for the velocity is defined at both ends such that 

\begin{equation} \label{eq:velBC}
    U\left(\chi=0,\tau\right)=0\qquad\text{and}\qquad U\left(\chi=1,\tau\right)=0
\end{equation}

\noindent which is independent of the value of $\eta$ \citep{Acharya_2021}. 

In order to define boundary conditions for $\alpha$ which are consistent with the ones defined for $U$, the conditions defined in equation (\ref{eq:velBC}) were applied on the non-dimensional momentum equation (\ref{eq:momentum_nondim}) to obtain $\partial p/\partial \chi = 0$. Then, the partial derivative of the non-dimensional pressure equation (\ref{eq:tubelaw_nondim}) was taken with respect to $\chi$ to obtain

\begin{equation}
\frac{\partial p}{\partial \chi}=\frac{\partial}{\partial \chi} \left(\frac{\alpha}{\theta}\right)- \eta \left( U\frac{\partial^2\alpha}{\partial\chi^2} + 
        2\frac{\partial\alpha}{\partial\chi}\frac{\partial U}{\partial\chi} + \alpha\frac{\partial^2 U}{\partial\chi^2} \right)=0.
\end{equation}

\noindent Since $\eta$ goes to $0$, a Neumann boundary condition for the non-dimensional area is obtained, such that

\begin{equation} \label{eq:areaBC_G}
    \left.\frac{\partial}{\partial \chi} \left(\frac{\alpha}{\theta}\right)\right|_{\chi=0,\tau}=0\qquad\mathrm{and}\qquad
    \left.\frac{\partial}{\partial \chi} \left(\frac{\alpha}{\theta}\right)\right|_{\chi=1,\tau}=0.
\end{equation}

\noindent The value of $\theta$ is constant at the boundary, hence the boundary condition for $\alpha$ is simplified to

\begin{equation} \label{eq:areaBC}T
    \left.\frac{\partial \alpha}{\partial \chi}\right|_{\chi=0,\tau}=0\qquad\mathrm{and}\qquad
    \left.\frac{\partial\alpha}{\partial\chi}\right|_{\chi=1,\tau}=0.
\end{equation}

\noindent Note that, as elaborated upon by \cite{Acharya_2021}, even in the case where $\eta$ is not equal to zero, the effects due to this damping at the boundary is negligible. 

\subsection{Numerical Implementation} \label{simulationBased}

Now that the dynamic problem is set, one needs to solve the system of equations (\ref{eq:continuity_nondim}) and (\ref{eq:momentum_nondim_final}) with the boundary conditions in equations (\ref{eq:velBC}) and (\ref{eq:areaBC}) to obtain the non-dimensional cross-sectional area ($\alpha$) and velocity ($U$). The initial conditions for the velocity and cross-sectional area are 

\begin{equation} \label{eq:velareaIC}
    U\left(\chi,\tau=0\right)=0  \qquad\text{and}
\qquad  \alpha\left(\chi,\tau=0\right)=\alpha_{\text{IC}}
\end{equation}

\noindent \citep{Acharya_2021}. A smoothing term $\epsilon\left(\alpha_{xx}\right)$ is added to the right-hand side of the non-dimensional continuity equation in (\ref{eq:continuity_nondim}) in order to reduce solving time by a faster convergence, such that

\begin{equation} \label{eq:continuity_nondim_stab}
    \frac{\partial\alpha}{\partial\tau}+\frac{\partial\left(\alpha U\right)}{\partial\chi} = \epsilon\left(\alpha_{xx}\right)
\end{equation}  

\noindent \citep{Acharya_2021}. This system was then solved using the MATLAB $\tt{pdepe}$ function. More details about the computational technique and validation of the solution is available in \cite{Acharya_2021}.

\section{Reduced-order Model} \label{reducedOrder}

\begin{figure*}[!htb]
    \centering{\fbox{\includegraphics[clip,width=0.7\textwidth]{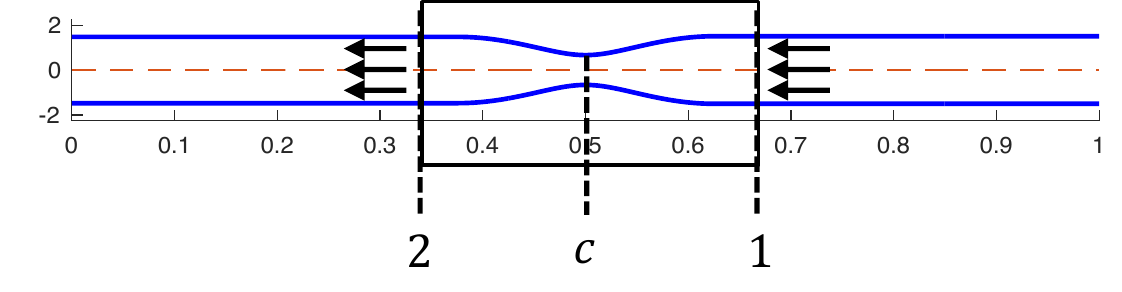}}}
    \caption{The geometry corresponding to the reduced-order model.}
    \label{fig:problem_geometry}
\end{figure*}

As an approximation, we also chose to study a simplified, reduced-order model of our dynamic problem by looking at the solution from the frame of reference of the traveling contraction wave at a specific point in time, $t=t_o$. The selected time instance is taken when the traveling contraction wave is located exactly at the center of the tube length. Half tube length is preferred so that the end effects are minimal and the approximation, discussed next, is a better one. In the chosen frame of reference and time instance, we assume that a fully developed flow goes from location 1 to location 2, as shown in figure~\ref{fig:problem_geometry}. Note that in the frame of reference of the lab, the peristaltic contraction wave travels at a constant speed $c_w$, in the opposite direction to the one shown in figure~\ref{fig:problem_geometry}, as it attempts to push the fluid towards the distal end of the tube (on the right). Therefore, throughout this paper, upstream refers to location $2$ and downstream refers to location $1$. 

For the fully developed flow configuration in figure~\ref{fig:problem_geometry}, the continuity and momentum are
\begin{equation} \label{eq:continuity_steady}
 \frac{\partial Q}{\partial x} = 0, \qquad \text{and}
\end{equation}

\begin{equation} \label{eqn: momentum conservation}
    \frac{\partial }{\partial x} \bigg[\frac{1}{2}\left(\frac{Q}{A}\right)^2 \bigg]=-\frac{\partial}{\partial x} \left(\frac{P}{\rho}\right)-\frac{f_D}{\rho},
\end{equation}

\noindent where Q is flow rate. The pressure equation (\ref{eqn: tube law}) is only a function of displacement given by

\begin{equation}
\frac{P(x)}{\rho}=\frac{K_{\scriptscriptstyle e}}{\rho}\left(\frac{A(x)}{A_{\scriptscriptstyle o}\theta(x)}-1\right).
\label{eqn: tube_law_disp}
\end{equation}

\noindent We can further simplify by plugging equation (\ref{eqn: tube_law_disp}) into equation (\ref{eqn: momentum conservation}) and integrating from point 1 to point 2 on figure~\ref{fig:problem_geometry} which yields 

\begin{equation}
\frac{\rho}{2{K_e}}\frac{Q^2}{{A_2}^2}+\bigg(\frac{A_2}{A_{\scriptscriptstyle o}\theta_2}-1\bigg)-\frac{\rho}{2{K_e}}\frac{Q^2}{{A_1}^2}-\bigg(\frac{A_1}{A_{\scriptscriptstyle o}\theta_1}-1\bigg)=-\int\frac{f_D}{K_e}dx,
\label{eqn: momentum conservation integral}
\end{equation}

\noindent where $\theta_1$ and $\theta_2$ are the relaxation factor at location 1 and 2, respectively. Note that the reference area at any location on the tube is defined as ${A_{\scriptscriptstyle o,k}(x_k,t)}={A_{\scriptscriptstyle o}\theta_k(x_k,t)}$. Therefore, ${A_{\scriptscriptstyle o}\theta_1}$ and ${A_{\scriptscriptstyle o}\theta_2}$ are the tube's reference areas at locations 1 and 2, denoted by ${A_{\scriptscriptstyle o,1}}$ and ${A_{\scriptscriptstyle o,2}}$, respectively. 

The non-dimensional form of equation (\ref{eqn: momentum conservation integral}) is

\begin{equation}
{A''_2}^3-\bigg(2\xi \bigg(\frac{\theta_2}{{\theta_1}}\bigg)^2+1-\psi'_\mu \bigg){A''_2}^2+2\xi=0,
\label{eqn:nonDim_momentum_conservation}
\end{equation}

\noindent where

\begin{equation}
{A''}_2=\frac{A'_2}{A'_1}, \qquad \xi=\frac{{\rho{Q^2}}{{A_o}\theta^3_1}}{4{K_e}{\theta^2_2}{A^3_1}} ,\qquad \text{and} \qquad {\psi}'_\mu=\frac{\int\frac{f_D}{K_e}dx}{A'_1},
\label{eq:A"_2 xi and psi'_mu}
\end{equation}

\noindent given that $A'_1={A_1}/{A_{\scriptscriptstyle o}\theta_1} $ and $ A'_2={A_2}/{A_{\scriptscriptstyle o}\theta_2}$. Lastly, we want to represent the cross-sectional area at the contraction (location $c$ in figure~\ref{fig:problem_geometry}) in terms of non-dimensional parameters. Since the contraction is located at point $c$, the cross-sectional area at $c$, denoted by $A_c$, is the smallest cross-sectional area of the tube.
When analyzing results we will use two different non-dimensionalizations of $A_c$ 

\begin{equation}
A''_c=\frac{A'_c}{A'_1}=\frac{A_c \theta_1}{A_1\theta_c } \qquad \text{and} \qquad A'''_c=\frac{A'_c}{A'_1}\theta_c =\frac{A_c}{A_1}{\theta_1},
\label{eq:non-dim parameters_Ac}
\end{equation}

\noindent where $\theta_c$ is the peristaltic contraction strength. The variable $A'_c$ is an intermediate non-dimensional variable for the contraction cross-sectional area, defined as $A'_c={A_c}/{A_{o,c} }={A_c}/{A_o\theta_c }$, where $A_{o,c}$ is the reference cross-sectional area of the tube at the strongest part of the contraction.

The non-dimensional parameters in equations (\ref{eq:A"_2 xi and psi'_mu}) and (\ref{eq:non-dim parameters_Ac}) are related to the non-dimensional parameters of the dynamic equations. Recall that $\alpha(\chi,\tau)=A(x,t)/A_o$, therefore

 \begin{equation}
A'_k=\frac{A_k}{A_{\scriptscriptstyle o}\theta_k} = \frac{\alpha(\chi_k,\tau_o)}{\theta_k},
\label{eqn:areaPrimei}
\end{equation}

\noindent where $k$ ($= 1$ or $2$) is the location on the tube. Moreover, the stiffness parameter $\xi$ of the reduced-order model can be expressed in terms of $\psi$, the stiffness parameter of the dynamic system, defined in section \ref{problemSetUp} as $\psi=K_e/(\rho c_w^2)$, such that

 \begin{equation}
\xi=\frac{\theta_1^3}{4\theta_2^2 \psi\alpha(\chi_1,\tau_o)}.
\label{eqn:xi_Psi}
\end{equation}

The viscous parameter $\beta$ maps onto the friction parameter ${\psi}'_\mu$. Since there is an integral in the definition of ${\psi}'_\mu$, its value depends on the shape of the elastic tube. In other words, there is no closed form, easy transformation between the two. However, it is possible to use the numerical solution of the 1D dynamic model and deduce the value of ${\psi}'_\mu$, which is related to the viscosity of the fluid. To solve for ${\psi}'_\mu$, equation (\ref{eqn:nonDim_momentum_conservation}) is rearranged such that 

\begin{equation}
\psi'_{\mu}=(1-A''_2)+2\xi\bigg[\bigg(\frac{\theta_2}{\theta_1}\bigg)^2-\frac{1}{A''_2}\bigg],
\label{eqn: friction_equation_full}
\end{equation}

\noindent Both non-dimensional parameters $A''_2$ and $\xi$ can be calculated from the solution of the 1D dynamic model as discussed above. If no relaxation is implemented then $\theta_1=\theta_2=1$ and
 
\begin{equation}
\psi'_{\mu}=(1-A''_2)\bigg[1-2\xi\frac{1-A''_2}{A''_2}\bigg].
\label{eqn: friction_equation_full_noRelax}
\end{equation}

\noindent Notice from the expression for $\xi$ in equation (\ref{eq:A"_2 xi and psi'_mu}), that the stiffness coefficient of the tube ($K_e$) is in the denominator, meaning that for a flow going through an elastic tube with very high stiffness, $\xi$ tends to zero. Hence, when the tube stiffness is sufficiently high, equation (\ref{eqn: friction_equation_full}) can be approximated by a linear relation between the friction parameter $\psi'_{\mu}$ and the non-dimensional upstream area $A''_2$, such that
 
\begin{equation}
\psi'_{\mu}\approx1-A''_2.
\label{eqn: friction_equation_linear_approx}
\end{equation}

\begin{figure*}
    \centering
    \begin{subfigure}[b]{0.475\textwidth}
        \centering
        \fbox{\includegraphics[trim=30 180 60 200,clip,width=\textwidth]{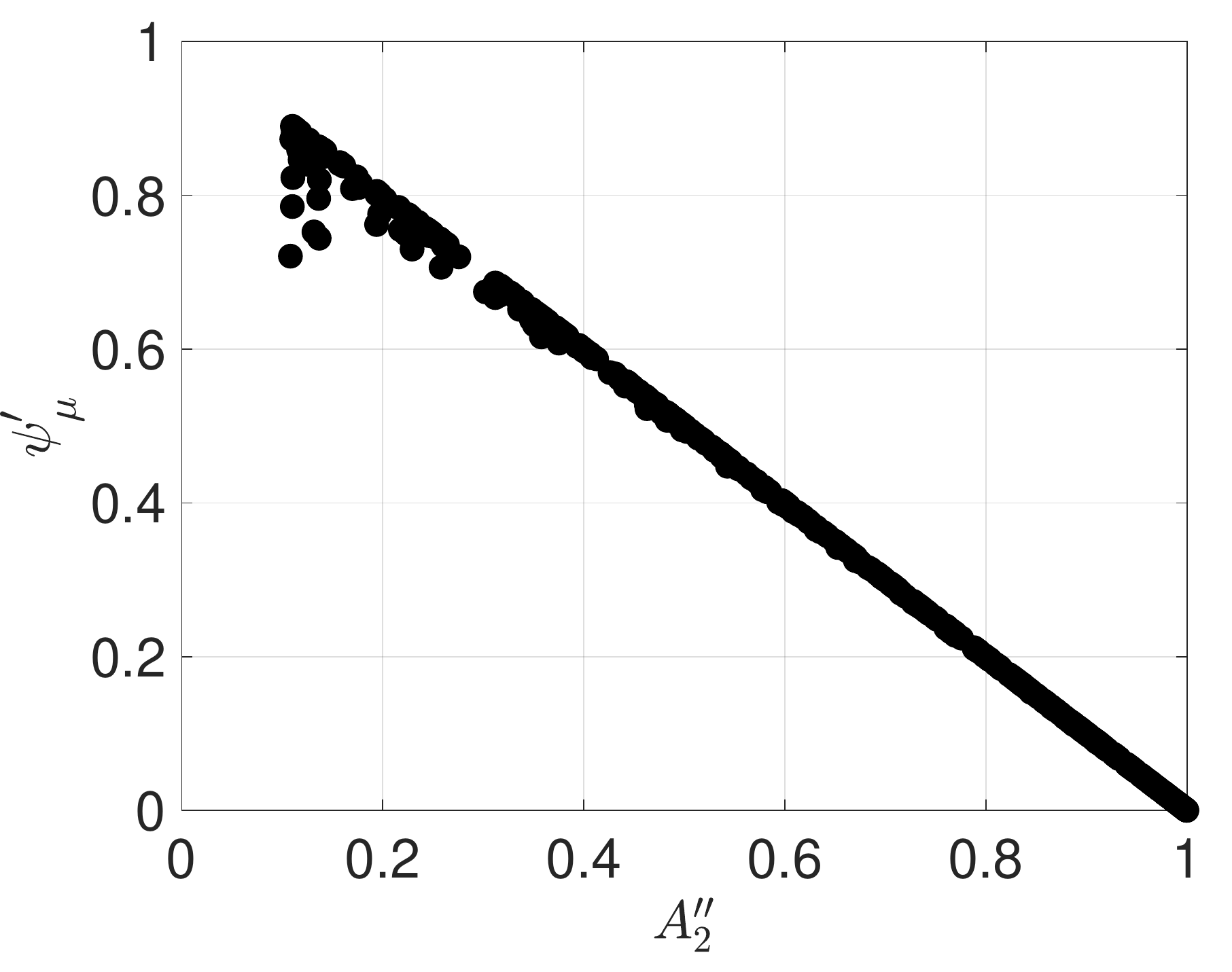}}
        \caption{No active relaxation, $\theta_1=1.0$}
        \label{fig:linRela_1}
    \end{subfigure}
    \hfill
    \begin{subfigure}[b]{0.475\textwidth}  
        \centering 
        \fbox{\includegraphics[trim=30 180 60 200,clip,width=\textwidth]{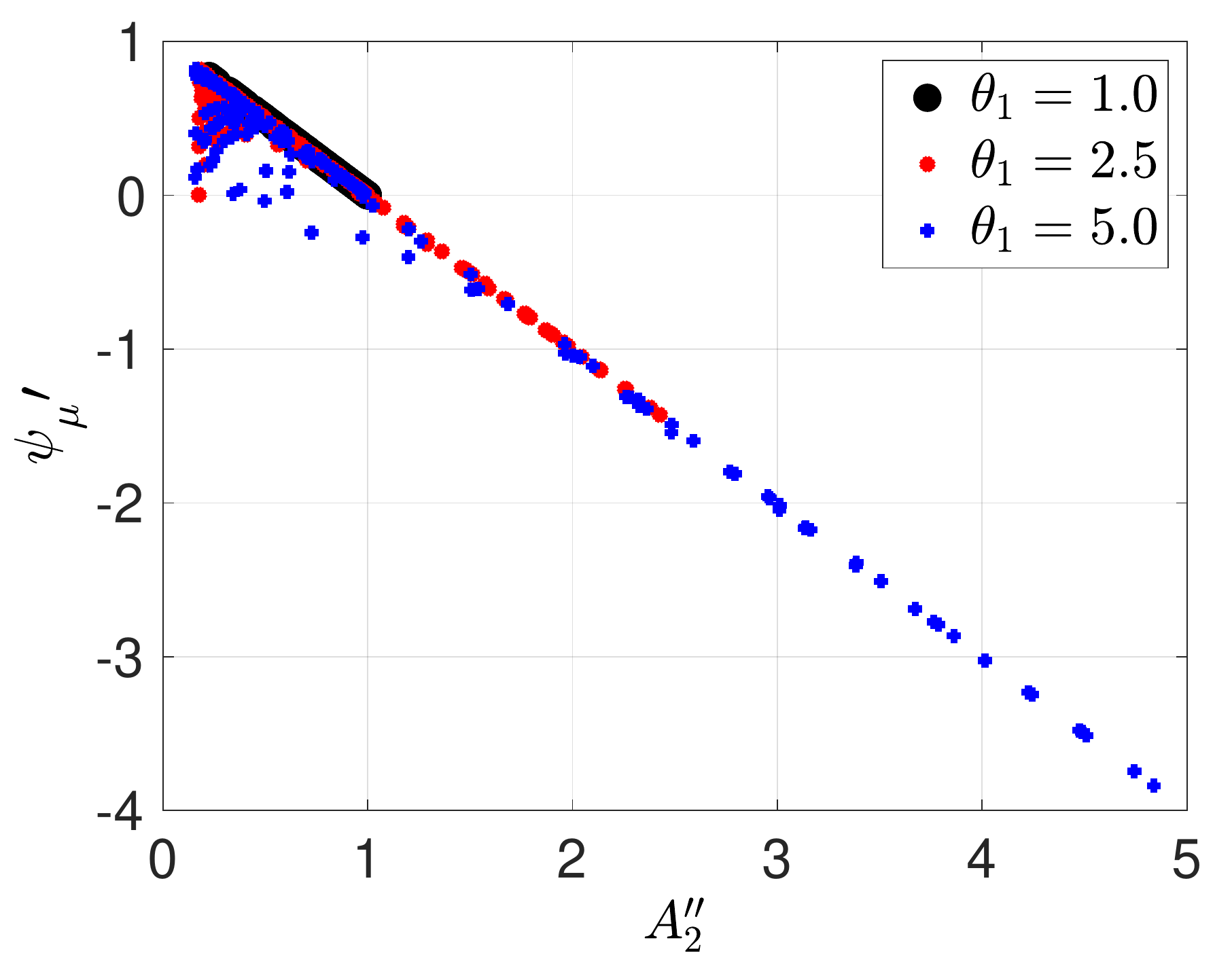}}
        \caption{Introduced active relaxation, $\theta_1=1.0, 2.5, 5.0$}
        \label{fig:linRela_all}
    \end{subfigure}
    \caption{Plotting the friction coefficient $\psi'_{\mu}$ and non-dimensional area parameter $A''_2$ to show their linear correlation.} 
    \label{fig:linear_relation_psiMue_theta1_1}
\end{figure*}

Using numerical solutions of the 1D dynamic model and equations (\ref{eq:A"_2 xi and psi'_mu}) and (\ref{eqn: friction_equation_full}), the non-dimensional area parameter $A''_2$ and the friction coefficient $\psi'_{\mu}$ were calculated and then plotted, as presented in figure~\ref{fig:linear_relation_psiMue_theta1_1}. Each point on the plots in the figure represents a single peristaltic contraction simulation, where each simulation encloses a different combination of the problem parameters, which include $\theta_c$, $\psi$, and $\beta$. Figure ~\ref{fig:linRela_1} displays the friction parameter as a function of non-dimensional cross-sectional area upstream (area 2) when no relaxation is implemented ($\theta_1=1$). Figure~\ref{fig:linRela_all} displays the same plot using three different data sets, each with a different relaxation strength, ($\theta_1 = 1.0$, $2.5$, and $5.0$), where the different strengths are distinguished by a different color and shape. As the figure shows, the simulation results are mostly consistent with the linear relation presented in equation (\ref{eqn: friction_equation_linear_approx}), with some outliers in each plot. The outliers, as expected, correspond to cases where the stiffness is not very high, introducing the nonlinearity in the relation. This linear relation is insightful since it provides a simple expression of friction in terms of cross-sectional area. Moreover, this relation is applicable to all three peristaltic regimes which presents a common property among all three resulting geometries. It is shown to be particularly useful in understanding and classifying the transition between regimes, as discussed in section~\ref{results}. 

\section{Results and Discussion} \label{results}

In the peristaltic regime investigation by \cite{Acharya_2021}, the peristaltic regime types were classified qualitatively, based on the resulting tube geometry, and these shapes were explained based on the problem's physical parameters. In this work, we aim to take this study one step forward by finding mathematical expressions to quantify and distinguish the different regimes numerically, and therefore, get a deeper understanding into the reasons behind their formation. Moreover, when relaxation is introduced, we find that the regime classification based solely on geometric shapes is not possible. Relaxation was not considered ($\theta_1 = 1.0$) in the prior work by \cite{Acharya_2021}. Therefore, here we seek a quantitative basis for regime classification. In addition, by defining the different peristaltic regimes analytically, we explain the transition between regimes 2 and 3 both quantitatively as well as qualitatively.

\subsection{Identification \& Mechanics of Different Peristaltic Regimes}\label{PumpingRegimes}

\begin{figure*}[!htb]
    \centering
    \begin{subfigure}[b]{0.460\textwidth}
        \centering
        \fbox{\includegraphics[trim=30 180 60 200,clip,width=\textwidth]{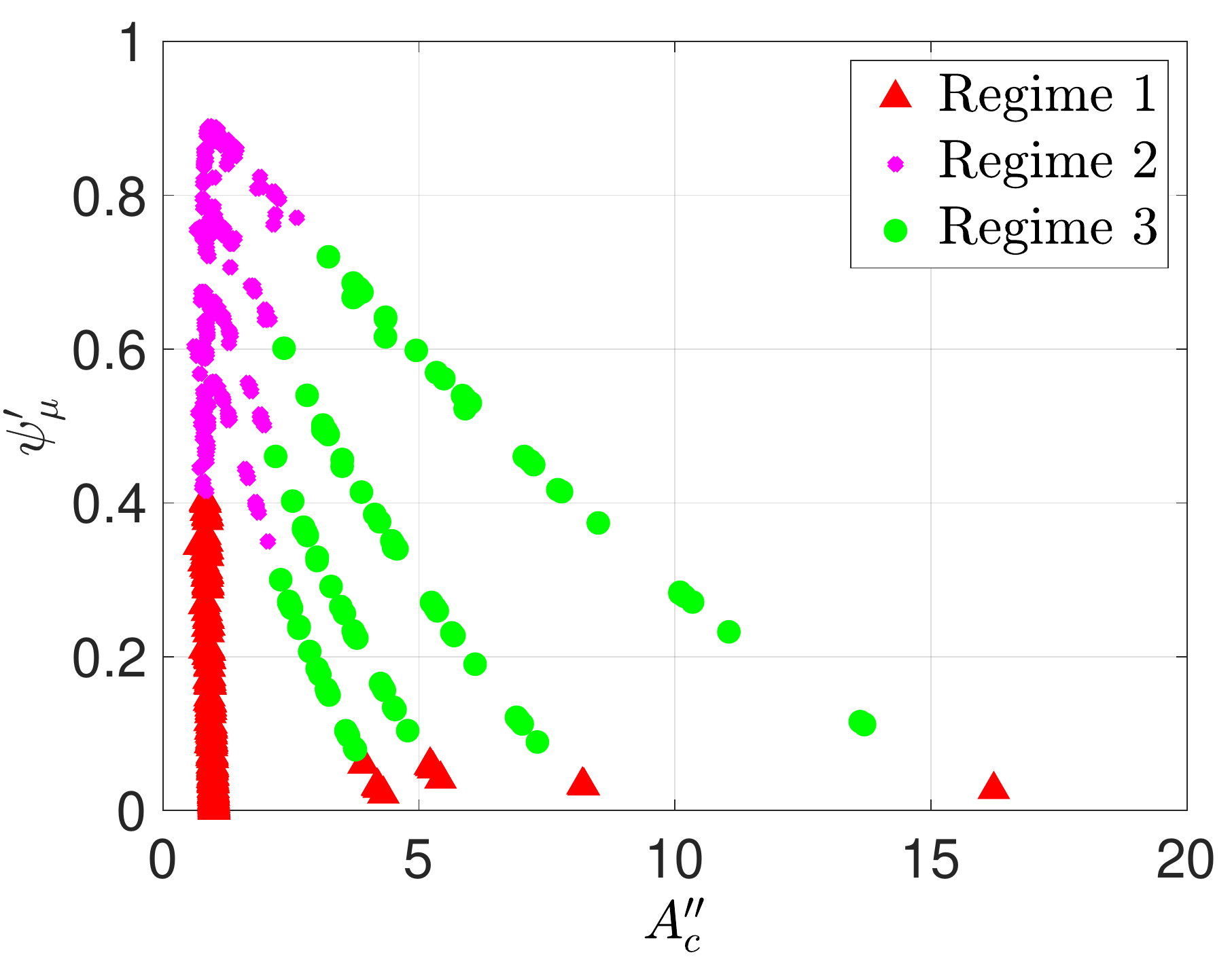}}
        \caption{No active relaxation, $\theta_1=1.0$}
        \label{fig:fric_vs_Adbpc_1}
    \end{subfigure}
    \hfill
    \begin{subfigure}[b]{0.490\textwidth}  
        \centering 
        \fbox{\includegraphics[clip,width=\textwidth]{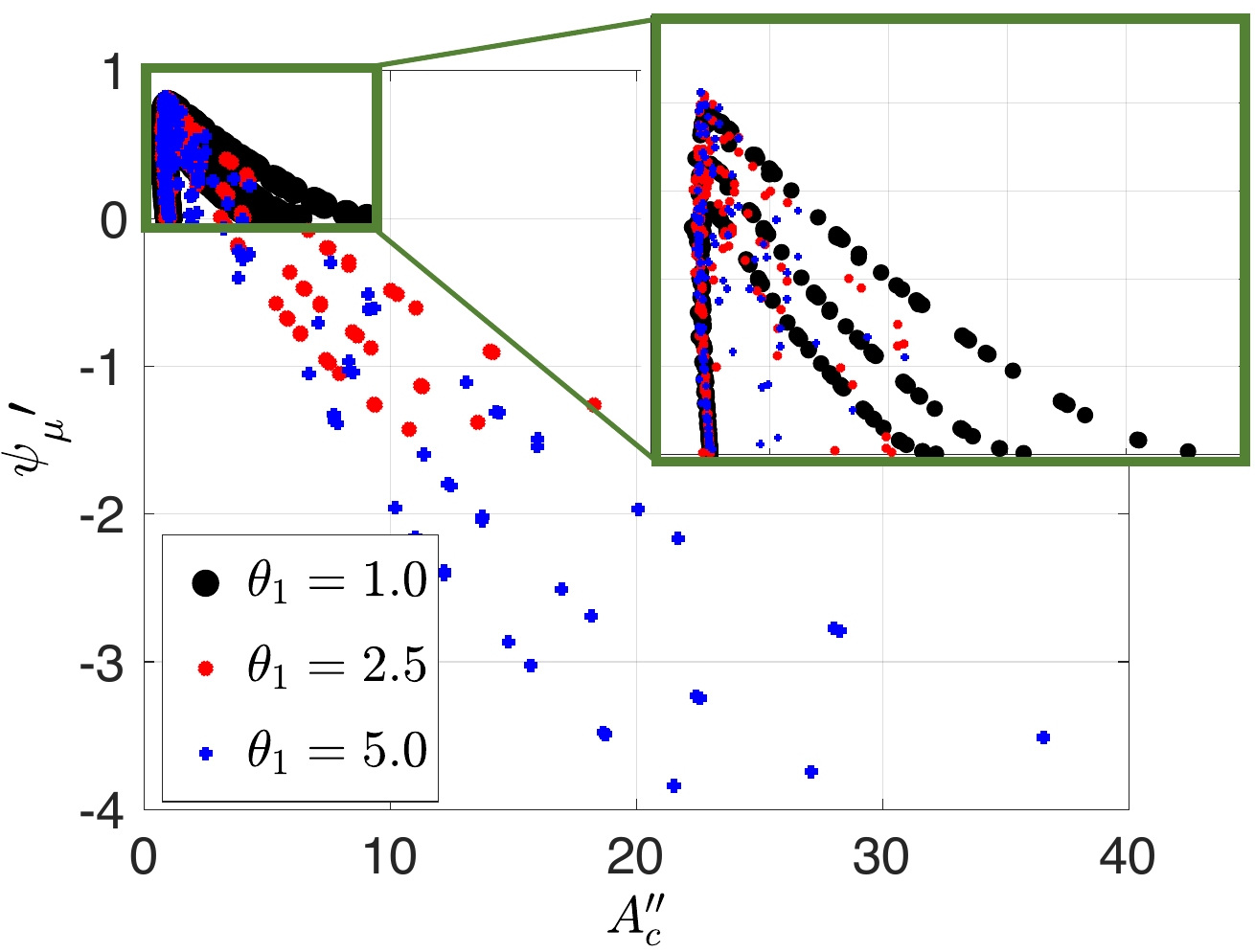}}
        \caption{Active relaxation, $\theta_1=1.0, 2.5, 5.0$}
        \label{fig:fric_vs_Adbpc_all}
    \end{subfigure}
    \caption{The friction parameter $\psi'_{\mu}$ plotted as a function of the non-dimensional contraction area parameter $A''_c$.} 
    \label{fig:fric_vs_Adbpc}
\end{figure*}

\begin{figure*}[!htb]
    \centering
    \begin{subfigure}[b]{0.462\textwidth}
        \centering
        \fbox{\includegraphics[trim=30 180 60 200,clip,width=\textwidth]{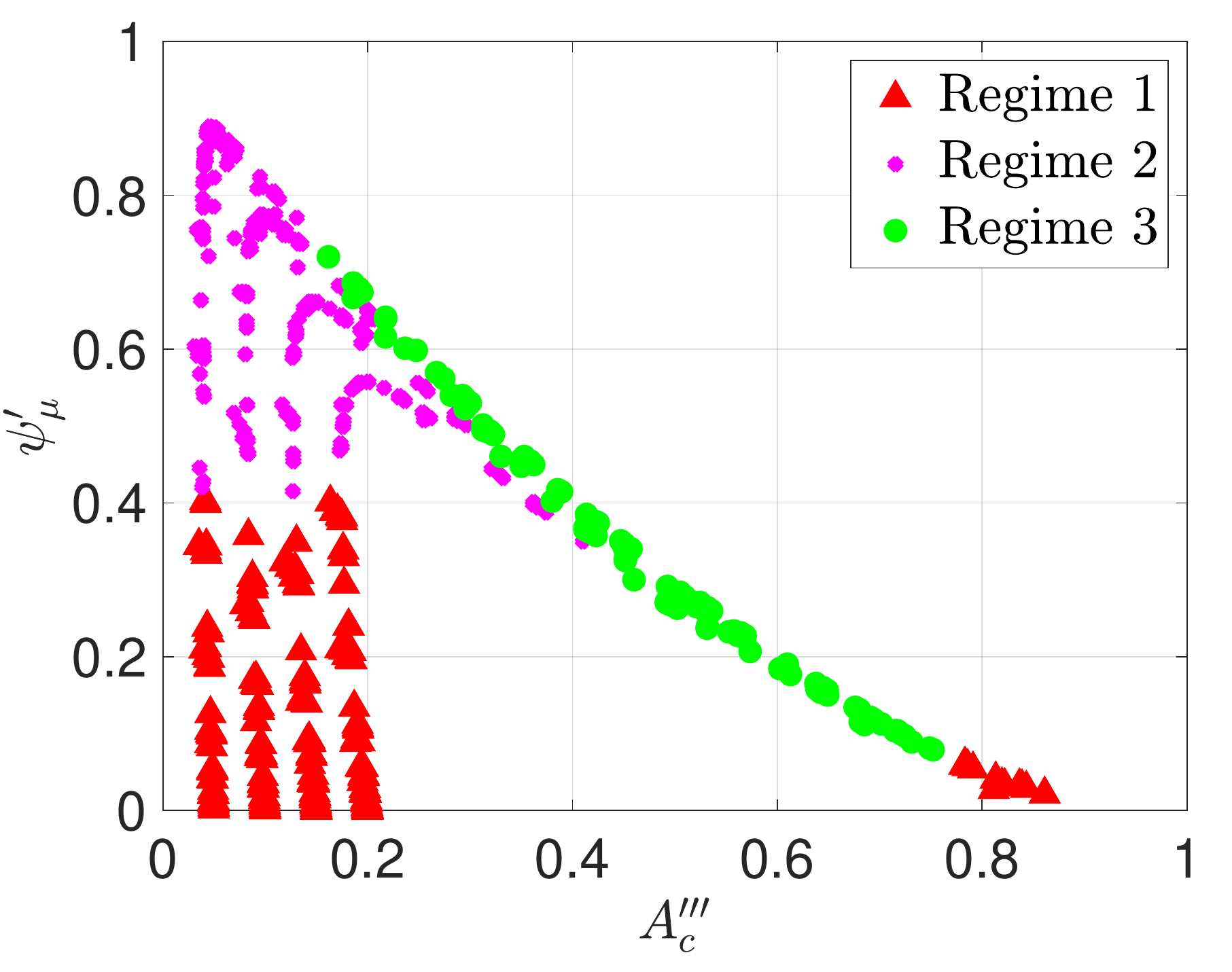}}
        \caption{No active relaxation, $\theta_1=1.0$}
        \label{fig:fric_vs_A3prc_1}
    \end{subfigure}
    \hfill
    \begin{subfigure}[b]{0.488\textwidth}  
        \centering 
        \fbox{\includegraphics[clip,width=\textwidth]{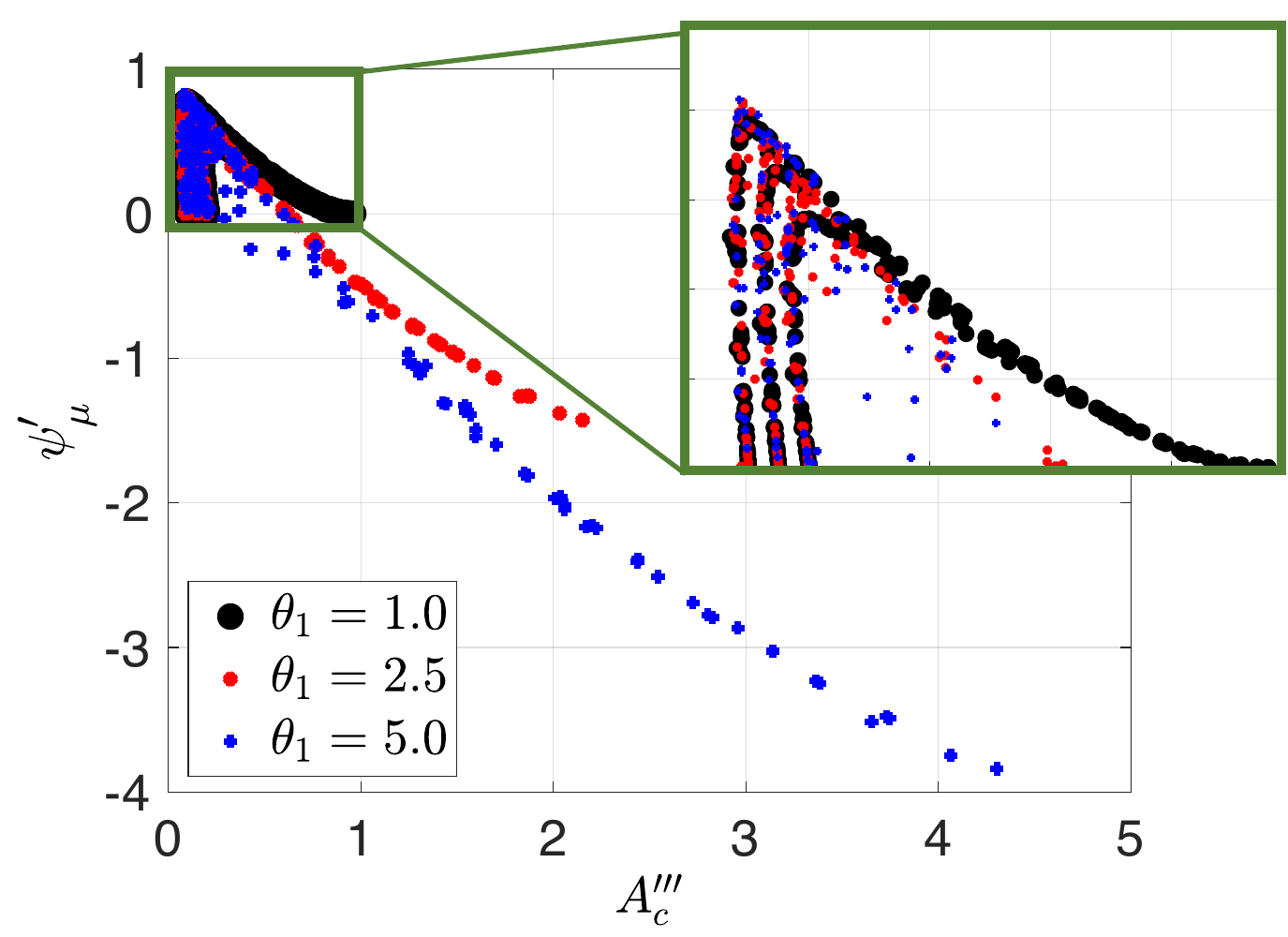}}
        \caption{introduced active relaxation, $\theta_1=1.0, 2.5, 5.0$}
        \label{fig:fric_vs_A3prc_all}
    \end{subfigure}
    \caption{The friction parameter $\psi'_{\mu}$ was plotted as a function of the non-dimensional contraction area parameter $A'''_c$.} 
    \label{fig:fric_vs_A3prime_c}
\end{figure*}

When looking at the geometries of the different peristaltic contractions obtained by the simulations' results without active relaxation (as shown in figure \ref{fig:regimes}), one can see that for peristaltic regimes 1 and 2, the cross-sectional area at location $c$ is much smaller than the cross-sectional areas at locations $1$ and $2$. In other words, the contraction is tight. On the other hand, in peristaltic regime $3$, the contraction cross-sectional area is not as tight, and the cross-sectional areas at locations $c$ and 2 seem to be of the same order of magnitude. Therefore, we hypothesize that the transition between peristaltic regimes 2 and 3 should be a process in which the contraction is relatively weak and the corresponding cross-sectional area opens. Moreover, \cite{Acharya_2021} concluded that the process of moving from peristaltic regime 1 to 3 is a direct result of an increase in viscosity. Hence, in the search for mathematical relations that characterize the different peristaltic regimes, we examine the role of friction in opening the contraction. To do so, we explore two different scalings' relations between friction and contraction area. The first, presented in figure ~\ref{fig:fric_vs_Adbpc}, displays the friction parameter $\psi'_{\mu}$ defined in equation (\ref{eqn: friction_equation_full}) as a function of the non-dimensional contraction area parameter $A''_c$ defined in equation (\ref{eq:non-dim parameters_Ac}). The second, presented in figure \ref{fig:fric_vs_A3prime_c}, displays the friction parameter $\psi'_{\mu}$ as a function of the non-dimensional contraction area parameter $A'''_c$ defined in equation (\ref{eq:non-dim parameters_Ac}).

Figure~\ref{fig:fric_vs_Adbpc_1} presents a plot of $\psi'_{\mu}$ (equation (\ref{eqn: friction_equation_full})) as a function of $A''_c$ (equation (\ref{eq:non-dim parameters_Ac})) for the simulations where no relaxation was present distal of the contraction $(\theta_1=1)$. In this figure, the peristaltic regime type of each case, as identified based on the shape of the tube, is marked in a different color and shape. By doing so, the plot reveals a clear separation between the three regimes, allowing us to determine the regime type based on the values of $\psi'_{\mu}$ and $A''_c$, without looking at the tube's geometry. As the figure shows, in this scaling, two clear patterns emerge. First, the simulation cases classified as regimes $1$ and $2$ fall on top of each other, creating a vertical line at a distinct value of $A''_c\approx1$. Second, the cases classified as regime $3$ form four distinct lines. Similarly, figure~\ref{fig:fric_vs_Adbpc_all} presents a plot of the friction parameter $\psi'_{\mu}$ as a function of $A''_c$ for three different data sets, each with a different relaxation strength, $\theta_1 = 1.0$, $2.5$, and $5.0$, where the different strengths are distinguished by a different color and shape. By looking at this plot, it is clear that it displays the same pattern as observed in figure~\ref{fig:fric_vs_Adbpc_1} (vertical line at $A''_c\approx1$ for regimes $1$ and $2$ and distinct lines for regime $3$). 

\begin{figure*}[!htb]

    \centering
    \begin{subfigure}[b]{0.7\textwidth}
        \centering
        \includegraphics[trim=75 350 60 350,clip,width=\textwidth]{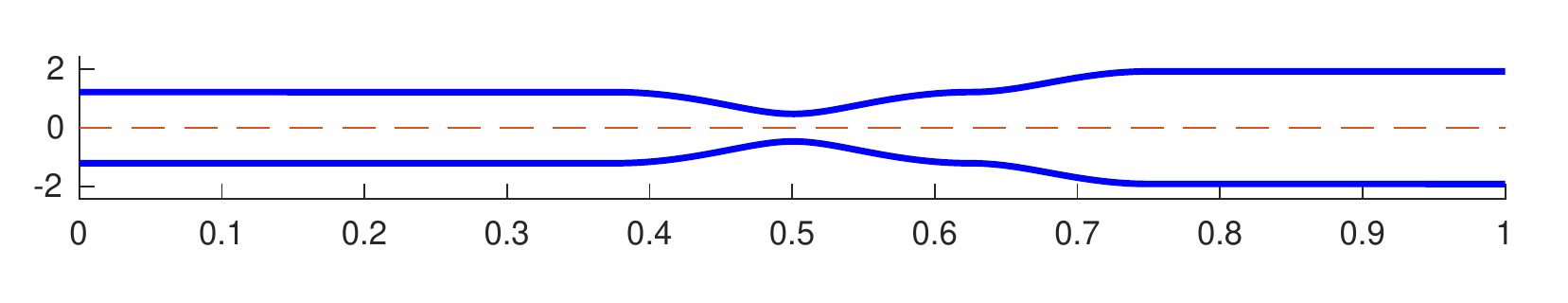}
        \caption{Regime 1}
        \label{fig:regime1_demo_R}
    \end{subfigure}
    \ 
    \begin{subfigure}[b]{0.7\textwidth}  
        \centering 
        \includegraphics[trim=75 350 60 350,clip,width=\textwidth]{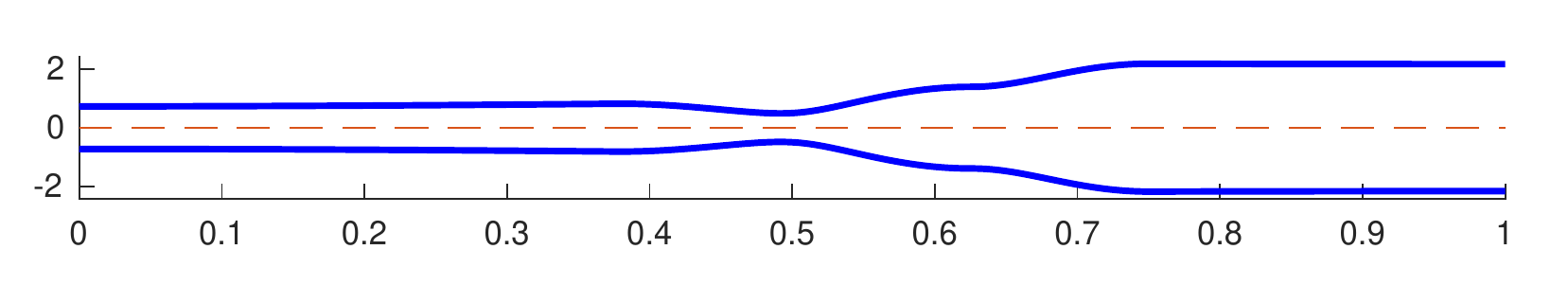}
        \caption{Regime $2$}
        \label{fig:regime2_demo_R}
    \end{subfigure}
    \ 
    \begin{subfigure}[b]{0.7\textwidth}   
        \centering 
        \includegraphics[trim=75 350 60 350,clip,width=\textwidth]{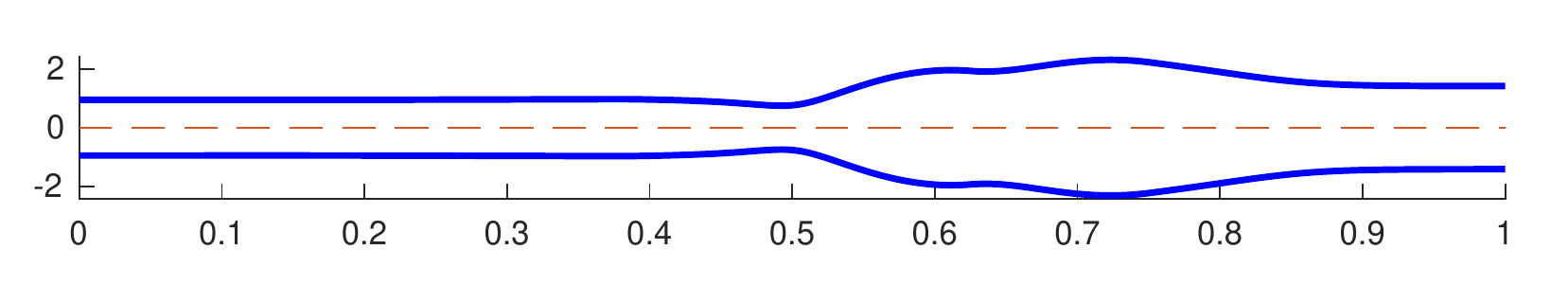}
        \caption{Regime 3}
        \label{fig:regime3_demo_R}
    \end{subfigure}
    \ 
    \caption{Tube geometry of the three peristaltic regimes in a FLIP device when there is relaxation distal of contraction. The separation into three regimes was done based on the observations similar to those in figure \ref{fig:fric_vs_Adbpc}.}
    \label{fig:regimes_R}
\end{figure*}

Noticing that all three data sets display a similar trend reveals a quantitative separation into peristaltic regimes that is independent of relaxation. Recall that when we implement active relaxation to the model, the resulting geometries of the tubes do not follow the previously defined classifications of regime since they have different shapes. Therefore, we did not have a way to characterize them into a certain regime. However, since the pattern from figure~\ref{fig:fric_vs_Adbpc_1} repeats in figure~\ref{fig:fric_vs_Adbpc_all}, we can conclude that contractile cycles that lie on the vertical $A''_c\approx1$  line are regimes 1 and 2, and the cases which lie right of this line are regime 3. Figure \ref{fig:regimes_R} displays the shapes of the three peristaltic regimes when relaxation is introduced as indicated by figure ~\ref{fig:fric_vs_Adbpc_all}.

Figure~\ref{fig:fric_vs_A3prc_1} presents a plot of the friction parameter $\psi'_{\mu}$ (equation (\ref{eqn: friction_equation_full})) as a function of $A'''_c$ (equation (\ref{eq:non-dim parameters_Ac})) for the simulations where no relaxation was present distal of contraction $(\theta_1=1)$. As in figure~\ref{fig:fric_vs_Adbpc_1}, the peristaltic regime type of each case, as identified based on the shape of the tube, is marked in a different color and shape. Different than the scaling used in figure \ref{fig:fric_vs_Adbpc}, in this scaling cases marked as peristaltic regime 3 merge into a single curve which is linear in leading order approximation. Moreover, the cases identified as peristaltic regimes 1 and 2 do not form a single vertical line but rather form four distinct vertical lines, each corresponds to a distinct value of $A'''_c = 0.05$, $0.10$, $0.15$, and $0.20$. Again, by identifying the regimes on this plot, we can see the separation between the three regimes, which provides us an additional way of determining the regime type quantitatively, this time based on the values of $\psi'_{\mu}$ and $A'''_c$.

Figure~\ref{fig:fric_vs_A3prc_all} presents a plot of $\psi'_{\mu}$ as a function of $A'''_c$ of three different data sets, each with a different relaxation strength, $\theta_1 = 1.0$, $2.5$, and $5.0$, where the different strengths are distinguished by a different color and shape. As the figure shows, three distinct vertical lines emerge at $A'''_c \approx 0.10$, $0.15$, and $0.20$, as well as a linear curve. These two trends are shared by all three data sets, which reveals a quantitative separation into peristaltic regimes that is independent of relaxation. Therefore, we can conclude that contractile cycles that fall on the vertical lines at $A'''_c \approx 0.10$, $0.15$, and $0.20$ are regimes 1 and 2, and the cases which lie on the linear curve are regime 3.

\subsubsection{Peristaltic Regimes 1 \& 2}\label{PumpingRegimes_1_2}

The observation that the cases classified as peristaltic regimes 1 and 2 form a single vertical line in figure \ref{fig:fric_vs_Adbpc} at $A''_c\approx1$ allows us to derive an analytic expression that characterizes these regimes and holds independent of the relaxation parameter. In equation~\ref{eq:non-dim parameters_Ac}, we defined $A''_c = {A'_c}/{A'_1}={A_c\theta_1}/{A_1\theta_c}$, which implies that

\begin{equation} \label{eq:Ac_doublePrime_expression}
    \frac{A_c\theta_1}{A_1\theta_c}\approx1,
\end{equation}

\noindent and therefore

\begin{equation} \label{eq:regime_1_and_2_Ac_A1_relation}
    A_c \approx \frac{A_1\theta_c}{\theta_1}.
\end{equation}

\noindent Equation (\ref{eq:regime_1_and_2_Ac_A1_relation}) reveals a direct relation between the cross-sectional area at the contraction and downstream cross-sectional area in the physical space for peristaltic regimes 1 and 2. The reason for this relation will be discussed later below.

The relation in (\ref{eq:regime_1_and_2_Ac_A1_relation}) is also observed in the scaling used in figure \ref{fig:fric_vs_A3prime_c}. The four values of $A'''_c$ which correspond to the four vertical lines of regime 1 and 2 in figure~\ref{fig:fric_vs_A3prc_1} are equivalent to the four contraction strengths, $\theta_c$, examined in this work. Therefore, for regimes 1 and 2, 

\begin{equation} \label{eq:regime_1_and_2_A3c_theta}
    A'''_c \approx \theta_c.
\end{equation}

\noindent Plugging the definition of $A'''_c$ from equation (\ref{eq:non-dim parameters_Ac}) into (\ref{eq:regime_1_and_2_A3c_theta}) results in the same relation presented in equation (\ref{eq:regime_1_and_2_Ac_A1_relation}). Note that for the simulations with active relaxation ($\theta_1>1$), the contraction strengths that were considered are $\theta_c = 0.10$, $0.15$, and $0.20$ and did not include $\theta_c = 0.05$, which explains why we only see three vertical lines in figure \ref{fig:fric_vs_A3prc_all}.

The relation presented in equation (\ref{eq:regime_1_and_2_Ac_A1_relation}) helps to explain why the contractile cycle cases classified as regimes $1$ and $2$ all fall on the same vertical line ($A''_c=1$) in figure~\ref{fig:fric_vs_Adbpc}. Imagine a case of an elastic tube that is empty and therefore not pressurized. The cross-sectional area of the unfilled tube is uniform and equal to the undeformed reference area (also known as rest area), $\theta A_o$. When applying a contraction of strength $\theta_c$ at location $c$ on the deformable, unpressurised tube, (simply squeezing the tube at point $c$), the area at location $c$ must satisfy $A_c =\theta_c A_o$. Next consider that the tube is starting to fill up with liquid. Consequently, the reference area will grow and the pressure will increase. As denoted in the linear tube law in equation (\ref{eqn: tube law}), the ratio $ {A}/{ \theta A_o}$ can be used to quantify the amount of filling because it helps us to understand the change in pressure. In the static case, the pressure is uniform throughout the tube length, and therefore

\begin{equation}
{P}={K_{\scriptscriptstyle e}}\left(\frac{A_1}{A_{\scriptscriptstyle o}\theta_1}-1\right)={K_{\scriptscriptstyle e}}\left(\frac{A_c}{A_{\scriptscriptstyle o}\theta_c}-1\right),
\label{eqn: static_Pressure}
\end{equation}

\noindent which can be simplified to 

\begin{equation}
\frac{A_1}{A_{\scriptscriptstyle o}\theta_1}=\frac{A_c}{A_{\scriptscriptstyle o}\theta_c},
\label{eqn: static_Pressure_1_c}
\end{equation}

\noindent and further rearranged to equation (\ref{eq:regime_1_and_2_Ac_A1_relation}). This shows that although the solutions considered in this analysis are of dynamic processes, contractile cycles classified as regimes $1$ and $2$ are very close to the static solution. In other words, the equation implies that the inertia effects of these cases are very small, so that they all fall on the same vertical line.

\begin{figure*}
    \centering{\fbox{\includegraphics[trim=0 180 0 200,clip,width=0.6\textwidth]{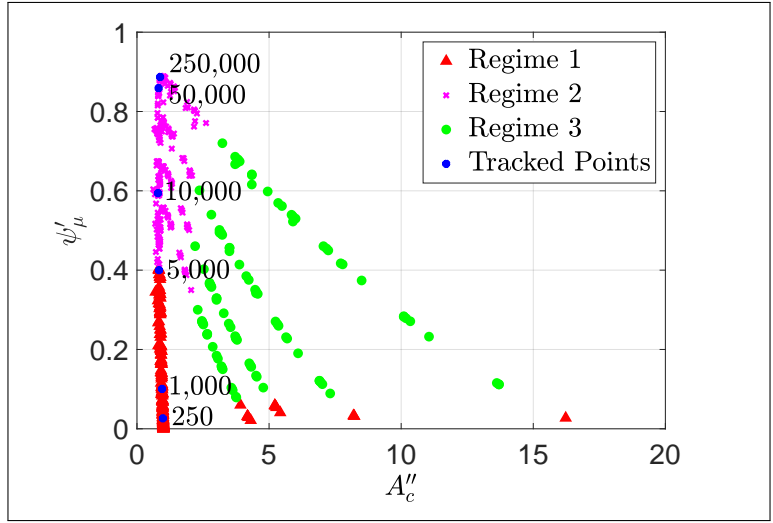}}}
    \caption{Plot of the friction parameter $\psi_\mu'$ as a function of $A''_c$ with six highlighted cases with the same $\psi$ and $\theta_c$ and changing $\beta$. The value of $\beta$ in each case is written on the plot. The plot shows how in peristaltic regimes 1 and 2, friction parameter increases as fluid viscosity parameter increases ($\beta$).}
    \label{fig:increaseBeta_regime_1_2}
\end{figure*}

One of the key characteristics of peristaltic regimes 1 and 2 is that the contraction remains tight. In other words, the cross-sectional area at location $c$ remains small and close to $\theta_c A_o$. This has been concluded by observing all the cases classified at peristaltic regimes 1 and 2. This is in contrast to the contraction in peristaltic regime 3, which opens, as discussed in section \ref{PumpingRegime_3}. Additionally, as expected, in regimes 1 and 2, when the fluid viscosity increases, the friction parameter (flow resistance) $\psi_\mu'$ increases. To confirm this, we track six cases classified as peristaltic regimes 1 and 2, all with the same wall stiffness ($\psi$) and contraction strength ($\theta_c$). The only parameter differentiating these cases is the value of $\beta$, the non-dimensional parameter which includes fluid viscosity. We observe that as the viscosity increases, the friction parameter increases. Figure \ref{fig:increaseBeta_regime_1_2} presents the plot in figure \ref{fig:fric_vs_Adbpc_1} with the six points identified, marked in blue and their corresponding $\beta$ values. The figure shows that even as the viscosity increases the contraction cross-sectional area parameter remains the same. Thus, the resistance to the flow increases at higher viscosity, which in turn causes an increase in $\psi_\mu'$. 

\subsubsection{Peristaltic Regime 3}\label{PumpingRegime_3}

Consider the line constructed for cases marked as regime 3 in both plots in figure~\ref{fig:fric_vs_A3prime_c}. To the leading order $\psi'_{\mu}$ goes as $1-A'''_c$, such that
\begin{equation} \label{eq:fric_approx_A3pc}
   {\psi'_{\mu}}\approx 1-A'''_c - C.
\end{equation}

\begin{figure*}
    \centering{\fbox{\includegraphics[width=0.6\textwidth]{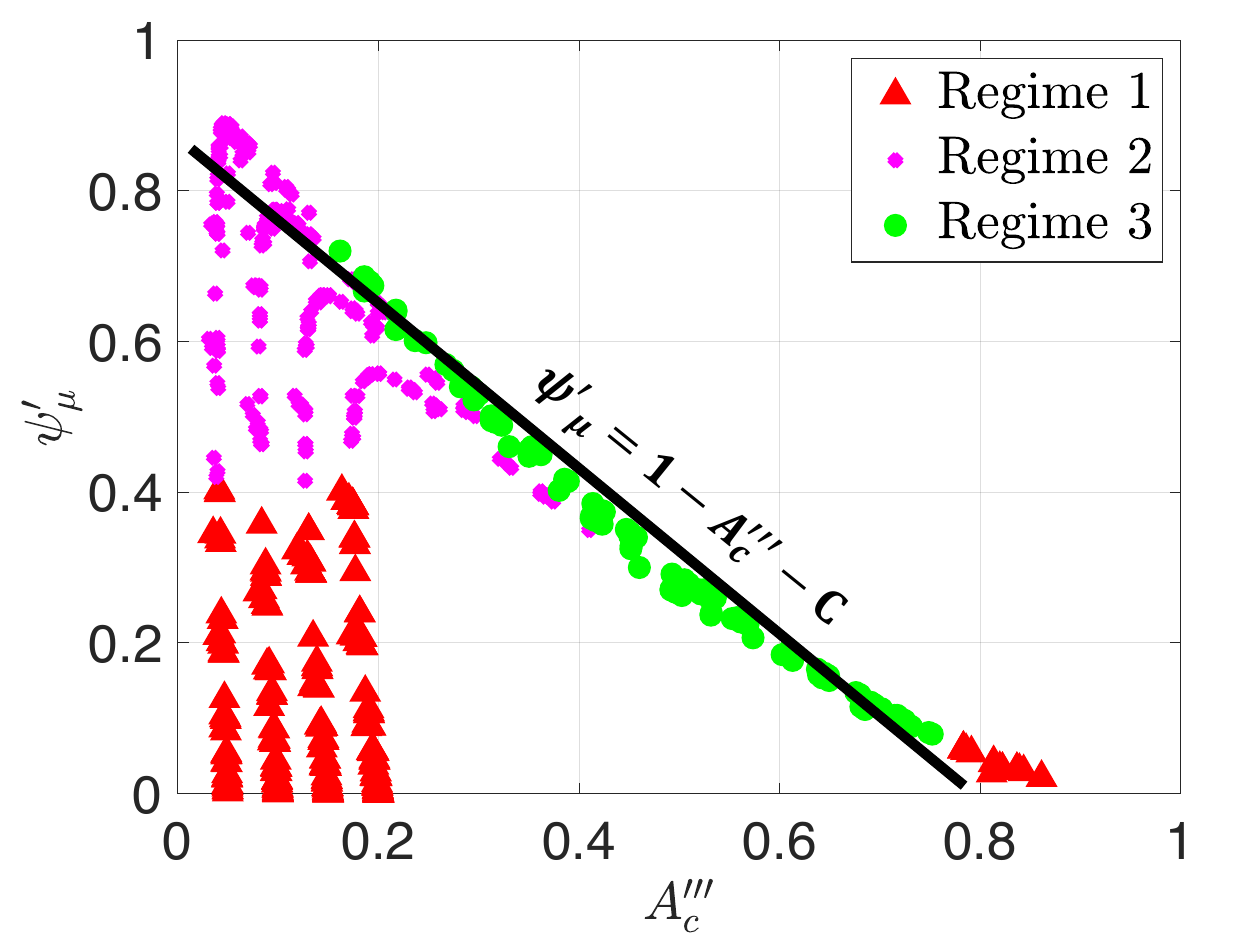}}}
    \caption{Plot of the friction parameter $\psi'_{\mu}$ as a function of $A'''_c$ alongside the line $\psi'_{\mu}=1-A'''_c-C$. This plot uses the output data from the simulations where there is no relaxation, $\theta_1=1.0$.}
    \label{fig:2_linear_Relat}
\end{figure*}

\noindent This simple linear relation expresses friction in terms of contraction area, that is unique to cases classified as peristaltic regime $3$. Most importantly, this expression, together with figure~\ref{fig:fric_vs_A3prime_c}, show that increasing the cross-sectional area at the contraction is related to decrease in friction. $C$ was determined to be approximately equal to $0.2$ based on figure~\ref{fig:2_linear_Relat}.

In order to make the expression in equation (\ref{eq:fric_approx_A3pc}) useful in differentiating between the peristaltic regimes, we want to find a relation between the area parameters that is unique to cases classified as regime 3. To do so, we utilize the derived linear relation in equation (\ref{eqn: friction_equation_linear_approx}), which applies to all three peristaltic regimes and equation \ref{eq:fric_approx_A3pc}, such that, for peristaltic regime 3,

\begin{equation} \label{eq:nonDim_A2_Ac}
   1-A'''_c-C\approx1-A''_2.
\end{equation}

\noindent Plugging in the definitions of $A'''_c$ and $A''_2$  from equations (\ref{eq:non-dim parameters_Ac}) and (\ref{eq:A"_2 xi and psi'_mu}), respectively, into equation (\ref{eq:nonDim_A2_Ac}), we obtain a relation between the contraction area and the area upstream in the physical space, that is unique to peristaltic regime 3, such that

\begin{equation} \label{eq:rregime_3_A2_Ac_relation_1}
     \frac{A_2} {\theta_2}\approx A_c + \frac{A_1} {\theta_1}C.
\end{equation}

\noindent Since in this work there is no active relaxation at location $2$, $\theta_2=1$ and

\begin{equation} \label{eq:regime_3_A2_Ac_direct}
   {A_2} \approx A_c + \frac{A_1} {\theta_1}C.
\end{equation}

\begin{figure*}
    \centering{\fbox{\includegraphics[trim=0 180 0 200,clip,width=0.6\textwidth]{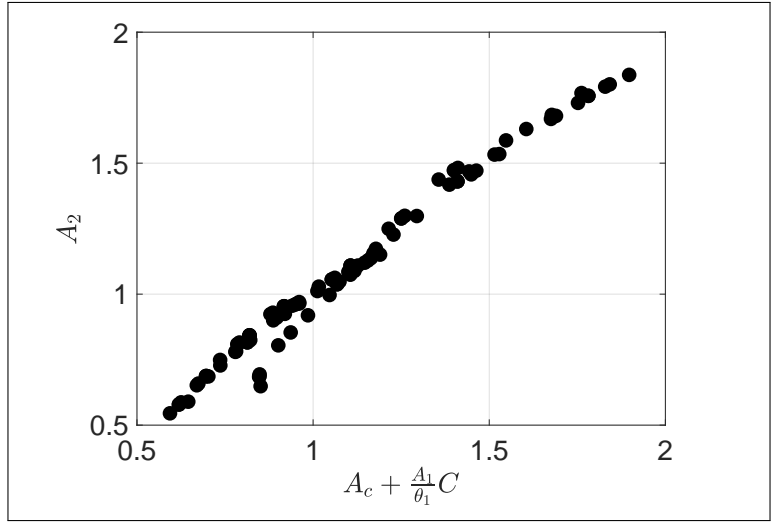}}}
    \caption{Plot of the actual cross-sectional area at location 2 as a function of the calculated cross-sectional area at location 2 from equation (\ref{eq:regime_3_A2_Ac_direct}) for cases classified as peristaltic regime 3. This plot shows an approximate linear line of slope 1, meaning that the model and data match well.}
    \label{fig:A2_Vs_A2Calc}
\end{figure*}

\noindent This relation tells us that in the cases classified as regime 3, the cross-sectional area at the location of the contraction is the same order as the area upstream, yet is still smaller. Note that we verified this relation by plotting $A_2$ as a function of $A_c + \frac{A_1} {\theta_1}C$ for the cases classified as regime 3, as presented in figure~\ref{fig:A2_Vs_A2Calc}. As this figure shows, the data formed an approximate line of slope $1$, which implies that there is almost an exact correlation between the two. This does not hold for cases classified as peristaltic regimes 1 and 2.

\begin{figure*}[!htb]
    \centering{\fbox{\includegraphics[trim=0 180 0 200,clip,width=0.6\textwidth]{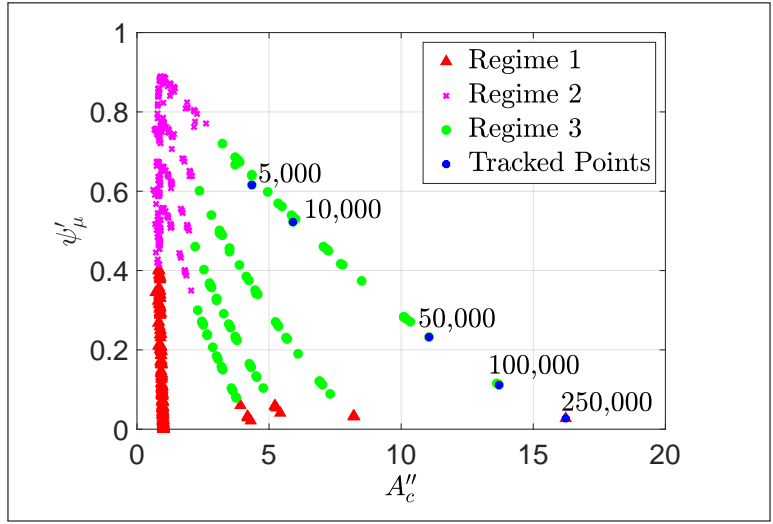}}}
    \caption{Plot of the friction parameter $\psi_\mu'$ as a function of $A''_c$ with five highlighted cases with the same $\psi$ and $\theta_c$ and changing $\beta$. The value of $\beta$ in each case is written on the plot. The plot shows how in peristaltic regimes 3, friction parameter ($\psi_\mu'$) decreases as fluid viscosity parameter ($\beta$) increases.}
    \label{fig:increaseBeta_regime_3}
\end{figure*}

Contrary to peristaltic regimes 1 and 2, one of the key characteristics of peristaltic regime 3 is that the cross-sectional area at the contraction opens, allowing fluid to go through. Specifically, what sets peristaltic regime 3 apart from peristaltic regimes 1 and 2 is that as the fluid viscosity (and consequently $\beta$) increases, the friction parameter $\psi_\mu'$ decreases. We concluded this based on simulation data. Figure \ref{fig:increaseBeta_regime_3} presents the plot in figure \ref{fig:fric_vs_Adbpc_1} with examined cases marked in blue and their corresponding $\beta$ values. The figure shows that as viscosity increases, the cross-sectional area of the contraction increases, which leads to a decrease in flow resistance, as also shown in equation \ref{eq:fric_approx_A3pc}. This observation makes physical sense. The contraction wave must travel, which cannot happen if fluid resistance distal of contraction is high. Therefore, while increasing fluid viscosity increases fluid resistance distal of contraction, the contraction compensates for that by opening, allowing fluid to flow upstream such that friction is reduced. 

\subsubsection{Transition Between Peristaltic Regimes 2 \& 3}\label{transition}

As previously discussed, understanding the transition process between peristaltic regimes and its causes can help in better understanding what causes a swallow to move from an effective to an ineffective region. In the previous sections, we revealed analytical relations between cross-sectional area parameters at different locations on the elastic tube. The expression in equation (\ref{eq:regime_1_and_2_Ac_A1_relation}) holds for peristaltic regimes $1$ and $2$ but not $3$, whereas the expression in equation (\ref{eq:regime_3_A2_Ac_direct}) holds only for peristaltic regime $3$. Therefore, we concluded that we can use these analytical expressions to differentiate between peristaltic regimes $1$ and $2$ to peristaltic regime $3$ quantitatively rather than qualitatively. Moreover, these relations provided reasoning beyond tube shape to the peristaltic regime classifications. In this section, we utilize these in order to explain the physical significance of the transition point and what causes the transition to occur. We derive mathematical expressions to identify and quantify the transition point between peristaltic regimes $2$ and $3$. Lastly, we identify the leading parameters that cause the transition between the two peristaltic regimes.

The key physical characteristics of peristaltic regimes 1 and 2 are tight contraction and that the friction parameter ($\psi_\mu'$) grows with the increase in fluid viscosity parameter ($\beta$). On the other hand, peristaltic regime 3 is characterized by opening of the contraction area and that the friction parameter decreases with an increase in fluid viscosity parameter. Therefore, the transition between peristaltic regime 2 and 3 must be the region in which the contraction area starts opening and the friction parameter is maximum. Cases classified as peristaltic regime 2, right before the transition, have high fluid resistance parameter but the contraction remains tight (i.e. with a small opening). As the fluid resistance keeps increasing, fluid starts accumulating distal of contraction (at location 1), until the resistance is so high, forcing the cross-sectional area at the contraction to open and allow fluid to go through. This opening reduces the overall friction to the flow, leading to regime 3 geometry. 

Since peristaltic regimes 2 and 3 have different analytical expressions associated with them (equations (\ref{eq:regime_1_and_2_Ac_A1_relation}) and (\ref{eq:regime_3_A2_Ac_direct}), respectively), the transition is identified at the point where both expressions hold. To find this expression, rearrange equation (\ref{eq:regime_1_and_2_Ac_A1_relation}) such that

\begin{equation} \label{eq:regime_1_and_2_Ac_A1_relation_rear}
     \frac{A_1} {\theta_1}= \frac{A_c} {\theta_c},
\end{equation}

\noindent then plug into equation (\ref{eq:regime_3_A2_Ac_direct}) to obtain

\begin{equation} \label{eq:transition}
   {A_2} = A_c \bigg(1+\frac{C} {\theta_c}\bigg).
\end{equation}

\noindent Equation (\ref{eq:transition}) represents the mathematical expression that only holds for contractile cycles in the transition between regimes 2 and 3. Since $\big(1+{C} /{\theta_c}\big)$ is a constant, the expression reveals that during transition, the cross-sectional area at location $2$ is proportional to the cross-sectional area at location $c$. 

\begin{figure*}
    \centering{\fbox{\includegraphics[width=0.65\textwidth]{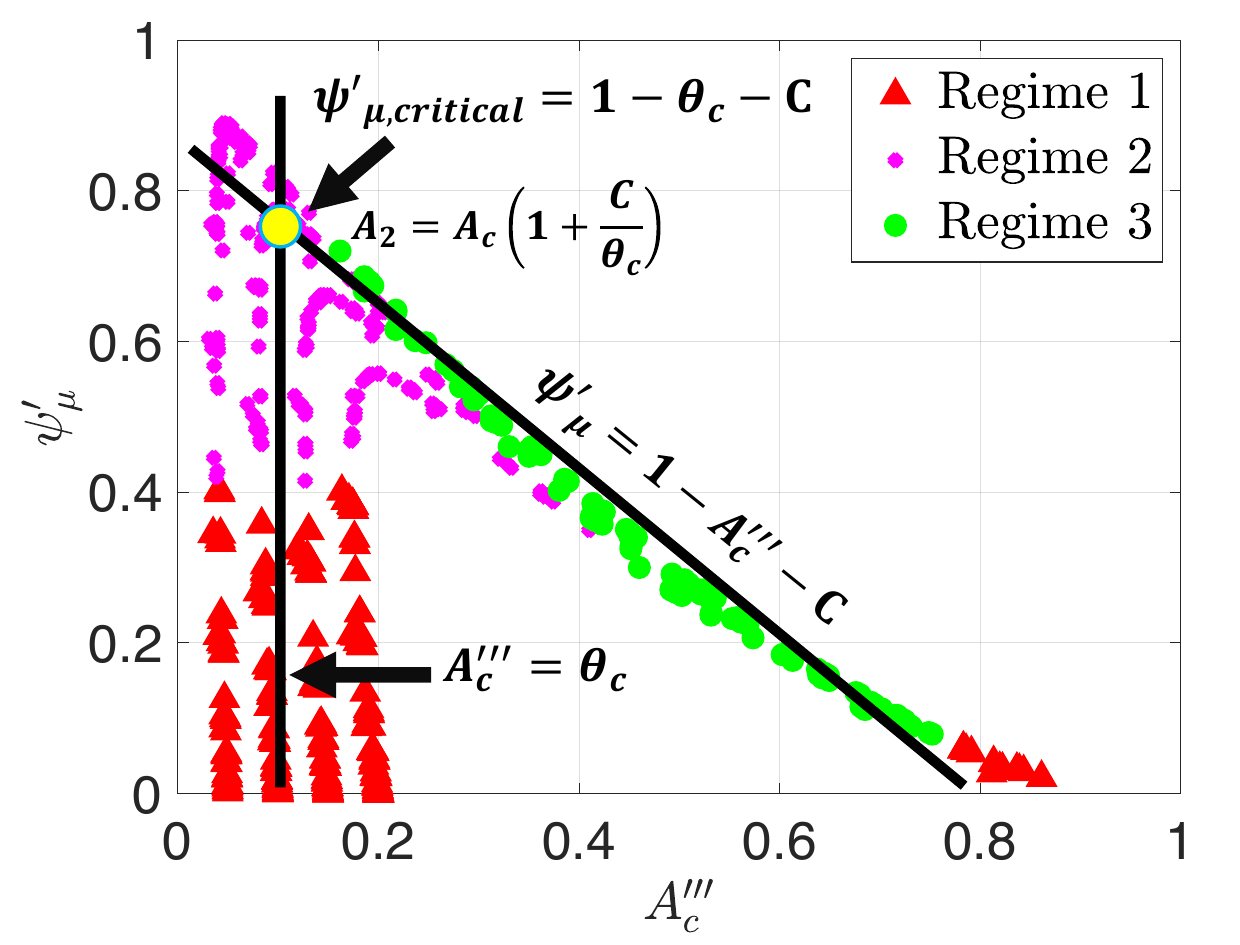}}}
    \caption{Plot of the friction parameter $\psi_\mu'$ as a function of $A'''_c$ with the two lines in equations (\ref{eq:regime_1_and_2_A3c_theta}) and (\ref{eq:fric_approx_A3pc}). }
    \label{fig:fric_vs_A3prime_c_transition}
\end{figure*}

Knowing the maximum friction value before the transition helps us to quantify the point of transition between peristaltic regime $2$ to $3$. Moreover, it allows us to identify the leading parameters that control the transition. The transition occurs at point ($A'''_c$, $\psi'_{\mu,critical}$) on the graphs in figure~\ref{fig:fric_vs_A3prime_c}, where $\psi'_{\mu,critical}$ is the maximum friction value before transitioning into peristaltic regime 3. The larger the value of $\psi'_{\mu,critical}$, the more resistance the contraction can withstand before opening and transitioning into regime 3. We can determine the location of this point on the graphs by finding the intersection of the lines in equations (\ref{eq:regime_1_and_2_A3c_theta}) and (\ref{eq:fric_approx_A3pc}) as shown in figure \ref{fig:fric_vs_A3prime_c_transition}, such that $A'''_c\approx\theta_c$ and

\begin{equation} \label{eq:transition_function}
   {\psi'_{\mu,critical}}\approx 1-\theta_c -C.
\end{equation}

\noindent Equation (\ref{eq:transition_function}) reveals that the point of transition between peristaltic regime 2 to 3 depends on the contraction strength. In addition, it tells us that the smaller the value of $\theta_c$, the larger the resistance needs to be in order to open the contraction. 

Notice from figure~\ref {fig:fric_vs_A3prc_1} that the transition into peristaltic regime $3$ occurs at a higher value of $\psi'_{\mu}$ when $A'''_c$ is small. In other words, the smaller $A'''_c$ is in regimes $1$ and $2$, the higher the resistance needs to be in order to open the area at location $c$. Since $A'''_c=\theta_c$ in regimes $1$ and $2$, this observation aligns with the analytical expression in equation (\ref{eq:transition_function}), where smaller contraction strength corresponds with larger $\psi'_{\mu,critical}$. The smaller the value of $\theta_c$, the tighter the contraction, and therefore, the harder it is to open. Hence, for a tighter contraction, more pressure needs to build up in the tube (caused by greater resistance to flow) in order for it to open. Consequently, a higher resistance parameter value is needed at transition.

Figure \ref{fig:fric_vs_A3prime_c} provides a view into the physical processes and factors that go into the transition between the peristaltic regimes. It clearly displays the relation between fluid viscosity, contraction strength, and the role they play in the transition process. In peristaltic regimes 1 and 2, $\psi'_{\mu}$ goes from 0 to some critical value where the transition occurs. This increase in fluid resistance is a result of the increase in fluid viscosity. In this regime, the contraction remains tight and effective in pushing the fluid forward. At the critical point, marked in figure \ref{fig:fric_vs_A3prime_c_transition}, the pressure build up due to high fluid resistance through the contraction is so high that the contraction is forced to open. This allows the fluid to go through the contraction more easily. The opening of the contraction (geometric change) decreases the friction parameter.

Wall stiffness also plays a role in the transition. As tube wall gets stiffer, it requires higher pressure to start opening the contraction area. High pressure builds up at the contraction if the fluid viscosity is greater. Thus, the parameter $\beta$ is greater for stiffer walls at the point of transition to regime 3. This is elaborated next. 

\begin{figure*}
    \centering
    \begin{subfigure}[b]{0.32\textwidth}
        \centering
        \fbox{\includegraphics[trim=30 180 60 200,clip,width=\textwidth]{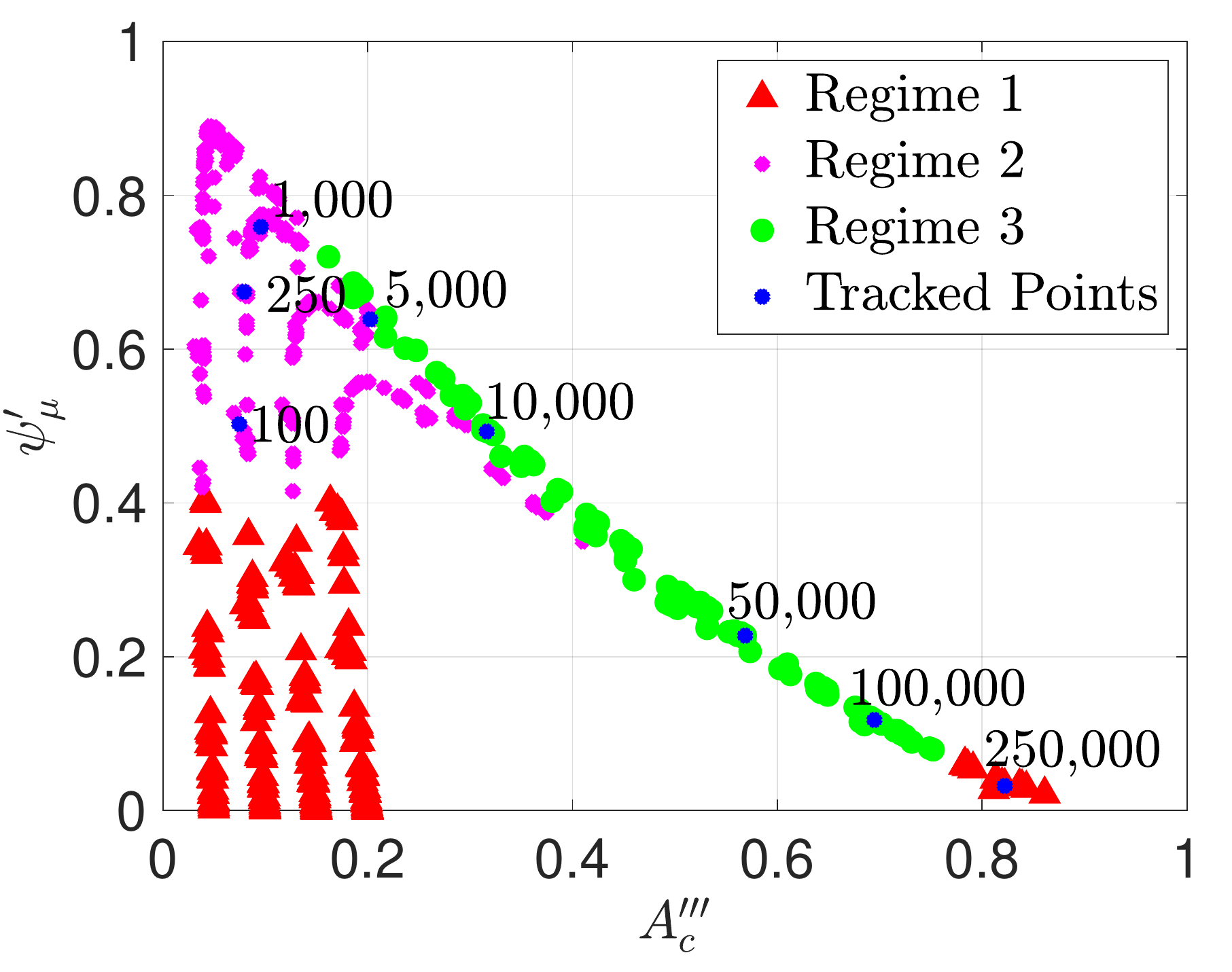}}
        \caption{$\psi=250$}
        \label{fig:psi250}
    \end{subfigure}
    \hfill
    \begin{subfigure}[b]{0.32\textwidth}  
        \centering 
        \fbox{\includegraphics[trim=30 180 60 200,clip,width=\textwidth]{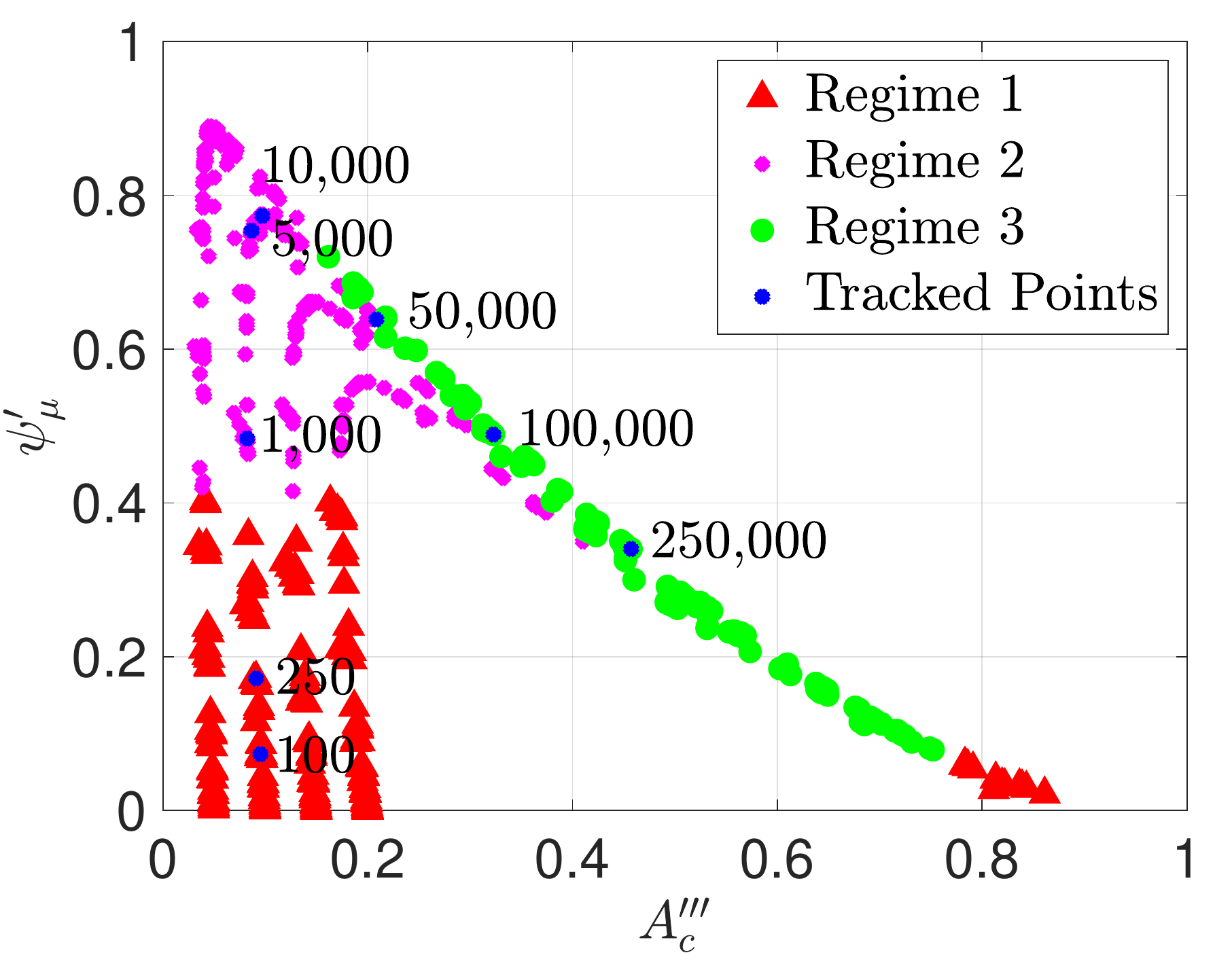}}
        \caption{$\psi=1,000$}
        \label{fig:psi1000}
    \end{subfigure}
     \hfill
    \begin{subfigure}[b]{0.32\textwidth}  
        \centering 
        \fbox{\includegraphics[trim=30 180 60 200,clip,width=\textwidth]{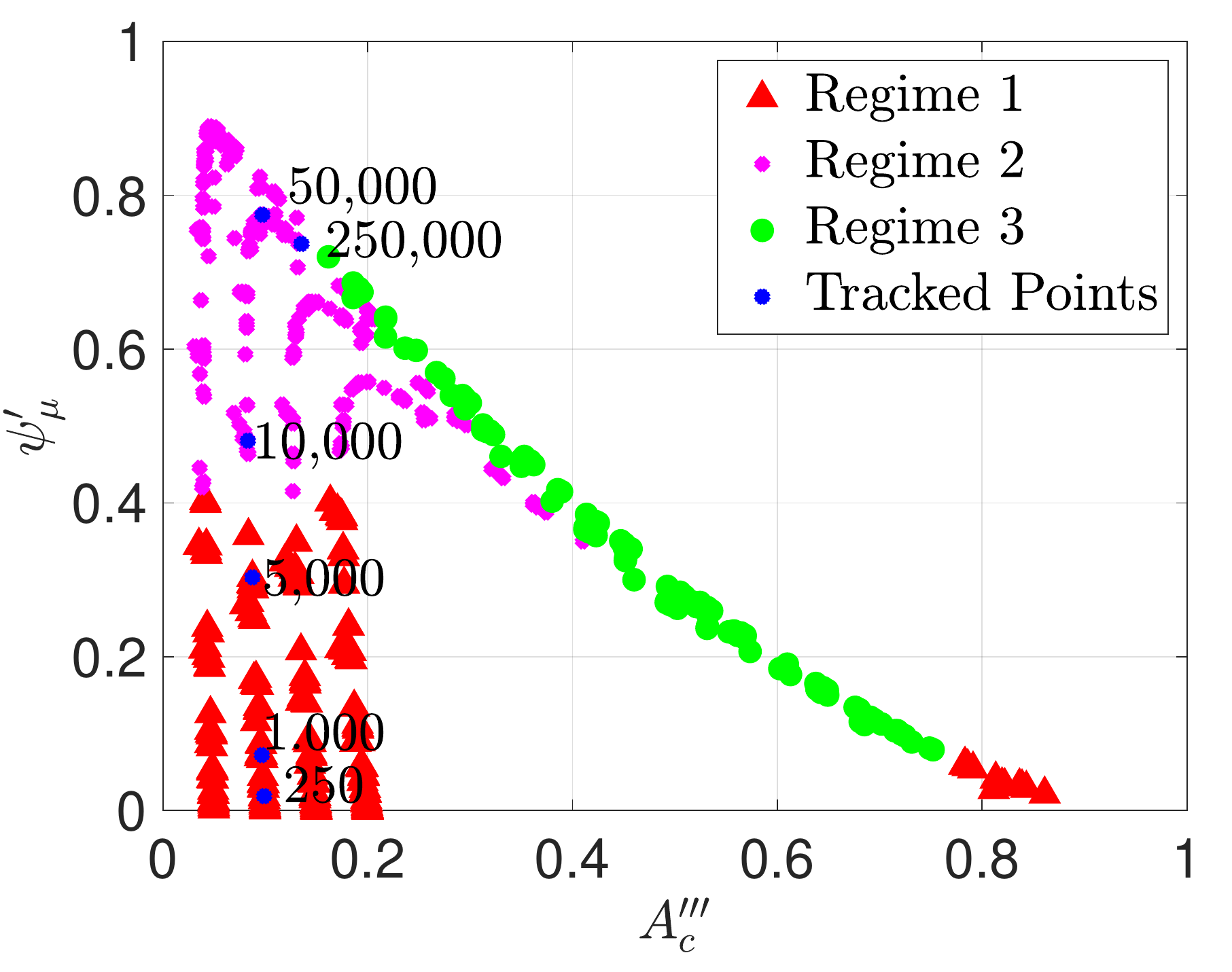}}
        \caption{$\psi=5,000$}
        \label{fig:psi5000}
    \end{subfigure}
    \caption{Three plots of the friction parameter $\psi_\mu'$ as a function of $A'''_c$, each with a different set of highlighted simulations. Each set shares contractile cycle simulations with the same contraction strength and wall stiffness, with changing fluid resistance parameter.} 
    \label{fig:trackBetaStiffness}
\end{figure*}

Figure \ref{fig:trackBetaStiffness} displays three $\psi_\mu'$ vs $A_c'''$ plots, each highlighting a different set of data in blue. Figure \ref{fig:psi250} highlights contractile cycle simulations with wall stiffness $\psi = 250$ and contraction strength $\theta_c = 0.15$, and changing fluid viscosity parameter, $\beta$. Figure \ref{fig:psi1000} shows similar results but for $\psi = 1,000$ and Figure \ref{fig:psi5000} for $\psi = 5,000$. The value of $\beta$ in each simulation is written next to the data point. Note that, since the value of $\theta_c$ is the same for all three sets, the value of the friction parameter $\psi_\mu'$ at the transition is the same, regardless of wall stiffness. This is not surprising since $\psi_\mu'$ captures the balance between passive elastic and viscous forces (see equation \ref{eq:A"_2 xi and psi'_mu}). Consequently, when stiffness is low, the transition occurs at a much lower value of fluid viscosity parameter ($\beta = 1,000$) than when the wall stiffness is high ($\beta = 50,000$).

\begin{figure*}
    \centering{\fbox{\includegraphics[clip,width=0.5\textwidth]{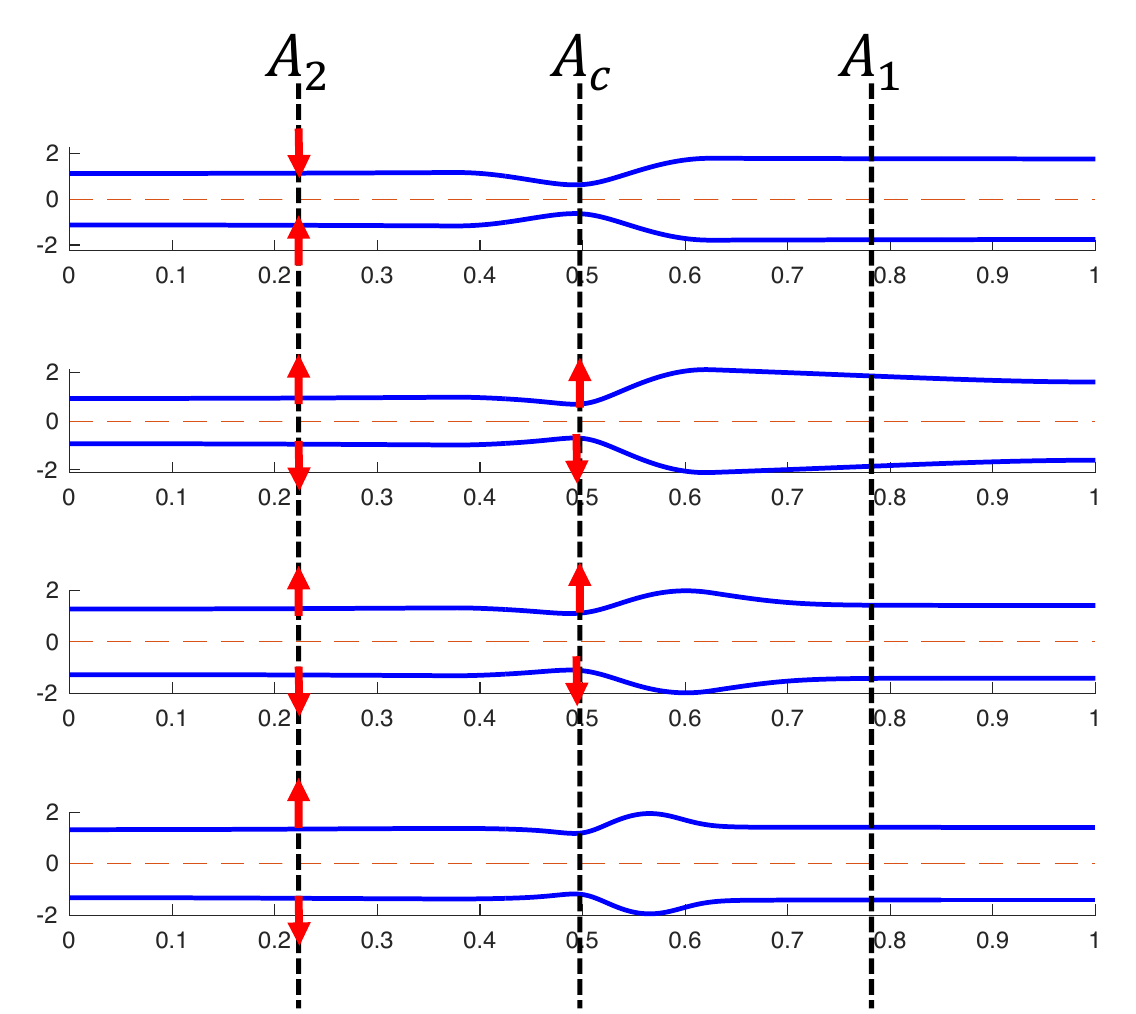}}}
    \caption{Tube geometry at transition}
    \label{fig:transition_physical}
\end{figure*}

In addition to expressing the transition from peristaltic regime 2 to peristaltic regime 3 analytically, we examine the shape of the tube during this process, which gives us additional insight into its physical progression. Figure~\ref{fig:transition_physical} presents the tube geometry of four different contractile cycles at one snapshot in time, taken when the peristaltic contraction wave is located at the center of the tube. The four cases are in consecutive order on the $\psi'_{\mu}$ vs. $A'''_c$ plot in figure \ref {fig:fric_vs_A3prc_1}. The tube shape at the top of the figure is classified as regime $2$, the bottom shape as regime $3$, and the middle two are transition cases. As the figure shows, the transition begins when the cross-sectional area $A_2$ starts decreasing until it reaches the same order of magnitude as the cross-sectional area at the contraction ($A_2\sim A_c$). At this point, as the figure shows, fluid has accumulated distal to the contraction. Beyond transition, the cross-sectional area at the contraction starts growing, allowing more fluid to go upstream and consequently increasing the cross-sectional area at location 2. Lastly, in the tube geometry at the bottom, the contraction is fully open, presenting regime 3 geometry.

\subsection{Application to Clinical Data}\label{clinical}

In order to examine the application of this study to clinical practice, we use FLIP readings of $56$ subjects, including both controls and patients, to calculate $\psi_\mu'$ (equation (\ref{eqn: friction_equation_full})) and $A_c''$ (equation (\ref{eq:non-dim parameters_Ac})). The FLIP data were collected between November 2012 and October 2018 at the Esophageal Center of Northwestern, using a 16-cm FLIP (EndoFLIP\textsuperscript{\tiny\textregistered} EF-322N; Medtronic, Inc, Shoreview, MN) \citep{AcharyaEsoWork2020,Carlson2021}. The patient population includes randomly chosen achalasia (n=9), GERD (n=13), systemic sclerosis (SSc; n=5), and eosinophilic esophagitis (EoE; n=5) patients. In addition, 24, randomly chosen, asymptomatic volunteers (“controls”) are included. Additional details on the clinical procedure and cohort selection are available in \citep{AcharyaEsoWork2020, Carlson2021,Lin2013,Carlson2016}. From each FLIP reading, we can extract several peristaltic cycles for each subject  \citep{Carlson2015}.  A single contractile cycle is identified by a transient decrease in the luminal diameter of at least 3 cm and distinct forward-moving contractions covering at least 6cm along the esophagus length \citep{AcharyaEsoWork2020,Carlson2021}. Since the analysis requires a clear peristaltic contraction wave, non-peristalsis subjects are eliminated. A lack of peristalsis is common among some esophageal disorders such as achalasia \citep{Aziz2016,Roman2011,Adler2001}. Therefore, a total of 103 different FLIP distention-induced contractions are used in this work, extracted from 34 subjects.

\begin{figure*}
    \centering{\fbox{\includegraphics[trim=60 320 60 350 ,clip,width=0.7\textwidth]{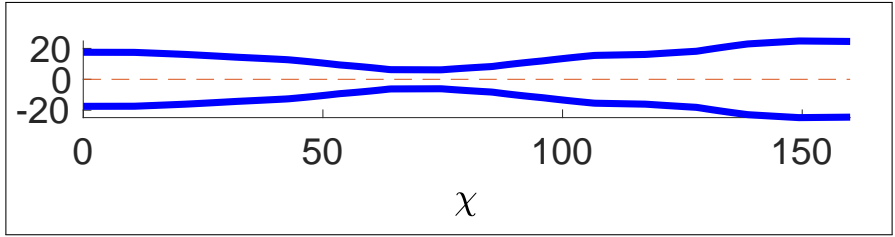}}}
    \caption{Esophagus wall shape of a control at a single time instance \citep{AcharyaEsoWork2020, Lin2013,Carlson2016}. The cross sectional data were captures using a FLIP device.}
    \label{fig:clinicalTubeArea}
\end{figure*}

Cross-sectional area values at locations 1, 2, and $c$ are direct FLIP readings, and $\theta_1$, $\theta_2$, and $\theta_c$ are calculated as proposed by \cite{Halder_2021}. These values are taken at a snapshot in time where the traveling contraction wave has traveled about half of the esophagus length, as shown in figure \ref{fig:clinicalTubeArea}. The fluid properties are density $\rho=1000\ \text{kg/m}^{3}$ and viscosity $\mu=0.001\ \text{Pa}\cdot\text{s}$ \citep{Kou2015ajpgi}. The wave speed $c_\omega=1.5-3\ \text{cm/s}$ as obtained by calculating the distance traveled by contraction wave over time \citep{Kou2015ajpgi,Li1994}. Lastly, the ratio $K_e/A_o$ is calculated using the tube law relation in equation (\ref{eqn: tube law}) \citep{Halder_2021}. Given this information, the parameters of interest is calculated. Note that $\psi_\mu'$ is calculated based on its definition in equation (\ref{eqn: friction_equation_full}). However, the non-dimensional stiffness parameter $\xi$ defined in equation (\ref{eq:A"_2 xi and psi'_mu}) tends to zero for all clinical cases and therefore, the approximate form of $\psi_\mu'$ presented in equation (\ref{eqn: friction_equation_linear_approx}) holds for clinical data.

\begin{figure*}
    \centering{\fbox{\includegraphics[clip,width=0.7\textwidth]{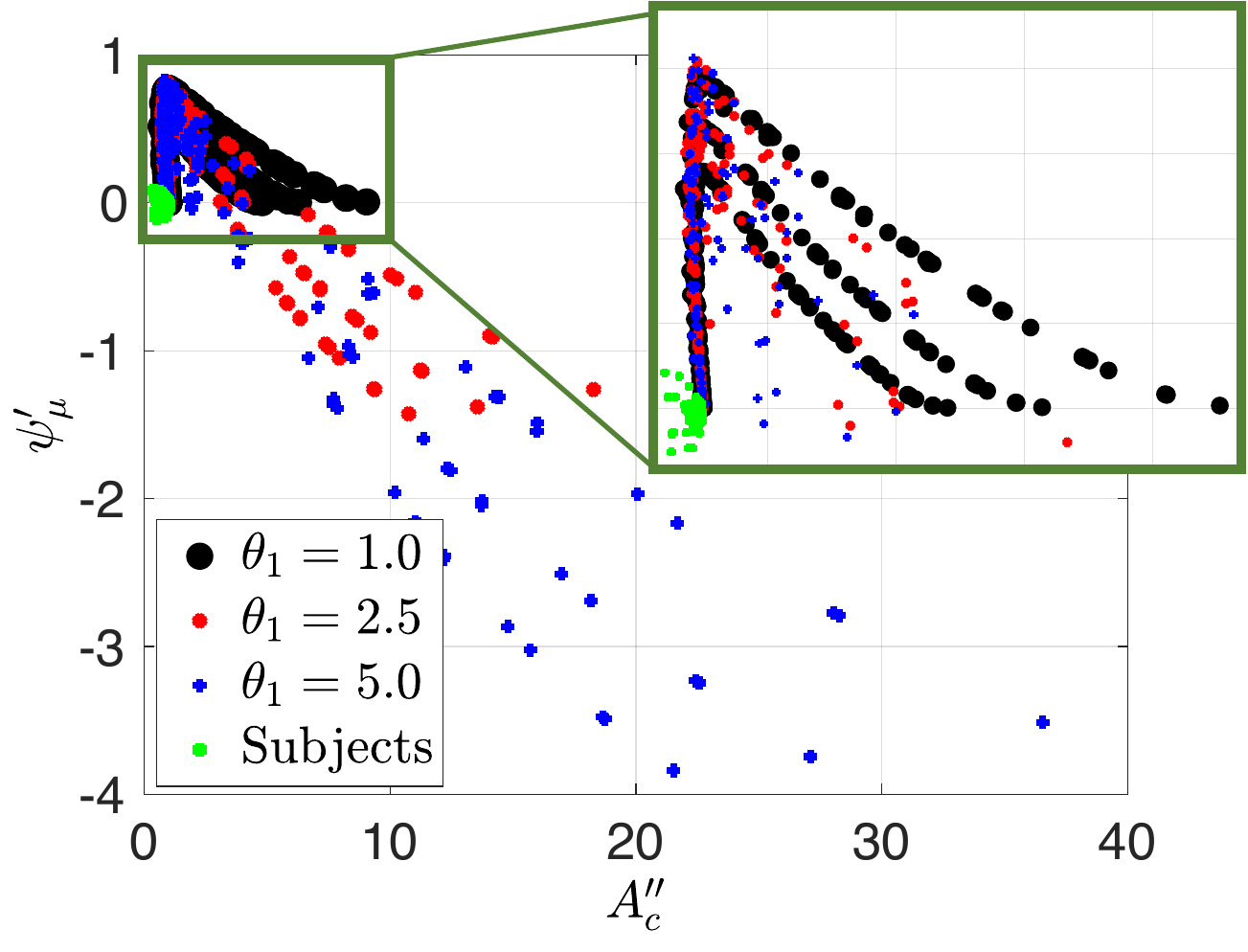}}}
    \caption{The friction parameter $\psi'_{\mu}$ plotted as a function of the non-dimensional contraction area parameter $A''_c$ for both simulation and clinical FLIP contractile cycles.}
    \label{fig:clinical_PsiMu_vs_Acdbp}
\end{figure*}

Figure \ref{fig:clinical_PsiMu_vs_Acdbp} presents a plot of the friction parameter $\psi_\mu'$ as a function of $A_c''$ for both simulation and clinical FLIP contractile cycles. Each point on the plot represents a single peristaltic contraction. As the figure shows, all the clinical cases lay on the vertical line, implying that they are in the regime 1 and 2 region. The outliers, which are clinical cases that lay left of the vertical line, all have small $\theta_c$ which makes them more error prone. Therefore, although they are slightly off of the vertical line, they display regime 1 and 2 geometry, and can be classified as such. Moreover, note that some of the clinical cases have relatively large $\theta_c$, such that the throat diameter looks big and can be mistaken for peristaltic regime 3. However, through this quantitative method of determining the regime, these cases are classified as regimes 1 and 2. 

No clinical case lies in the regime 3 region. Therefore, we concluded that although peristaltic flows have three regimes of pumping in general, physiological flows display only regimes 1 and 2. An interesting question is, why are physiological cases so far removed from regime 3? Two hypotheses are worth considering. The first hypothesis concerns the evolutionary disadvantage of peristaltic regime 3. Regime 3 implies hardship of transport due to the fact that the contraction opens and is no longer effective in pushing fluid forward. Animals cannot survive without an effective peristalsis which transports swallowed material from the mouth to the stomach. Therefore, peristaltic regime 3 does not align with the healthy function and core purpose of the esophagus, which is to transport swallowed material through a peristaltic contraction of the muscle. 

The second hypothesis concerns fluid properties. As discussed in section \ref{transition}, moving into the regime 3 region requires an increase in fluid viscosity. Therefore, in order to see a physiological example in regime 3 range, one needs to consider more viscous fluids. However, what fluid viscosity causes a healthy control to tip into regime 3? To answer this question, we first obtained that for clinical data $\psi\sim O(10^3)$, $\theta_c\sim O(10^{-1})$, and $\beta\sim O(1)$. Figure \ref{fig:psi1000} follows cases with similar $\psi$ and $\theta_c$ values. As the figure shows, given these values of $\psi$ and $\theta_c$, the transition into peristaltic regime 3 takes place at $\beta\approx 50,000$. Based on the definition of $\beta=8\pi L\mu/vA_o\rho$, the kinematic viscosity of the fluid needs to be about $25,000-50,000$ times that of water (i.e. $25,000-50,000\ \text{cP}$) for regime 3 to occur. The exact viscosity of chewed material is hard to determine due to the large variety of mastication and eating patterns. Therefore, studies often use texture of a puree or pudding \citep{Matsuo2013,Glassburn1998,Clave2008,Nicosia2012,Clave2006,Repin2018}, which have kinematic viscosity of $\nu=3,000-4,000\ \text{cP}$ \citep{Clave2008,Qasem2017,Lim2006,Kay2017}; much lower than $25,000-50,000\ \text{cP}$ \citep{Nicosia2012,Repin2018,Lim2006}. Hence, it is highly unlikely to obtain peristaltic regime 3 in subjects who display intact peristalsis. Nevertheless, examining clinical data collected using a higher viscosity fluids or solid food might help test this analysis further in future studies. In case regime 3 is observed for solid foods, it could act as an early indicator for insufficient peristalsis.

Note that some disease and disorder groups such as achalasia involve lack of esophageal peristalsis to some extent \citep{Aziz2016,Roman2011,Adler2001}. Since obtaining $\psi_\mu'$ (equation (\ref{eqn: friction_equation_full})) and $A_c''$ (equation \ref{eq:non-dim parameters_Ac})) requires a visible traveling peristaltic contraction, subjects without an identified contraction are classified as non-peristaltic cases and therefore are not included in the subject cohort above. Recall that regime 3 is characterized by opening of the contraction. Hence, it is plausible that in some cases, the peristaltic contraction is so weak that it opens easily, which appear in FLIP reading as nonexistent. As figure \ref{fig:fric_vs_A3prime_c} suggests, distention-induced peristalsis with contraction strength ($\theta_c$) larger than about $0.7-0.8$ always result in peristaltic regime 3 geometry, independent of the fluid resistance or wall parameters. The largest value of the contraction strength recorded in peristaltic cases is $0.5$. 

\section{Concluding Remarks} \label{conclusion}

\cite{Acharya_2021} examined the different elements of peristaltic flow by simulating a 1D peristaltic flow through an elastic tube closed on both ends. They qualitatively identified three modes of peristaltic geometries presented in figure \ref{fig:regimes} \citep{Acharya_2021}. In this work, we aimed to take \cite{Acharya_2021} results a step forward and quantify the different peristaltic geometries. Doing so allows us to differentiate the peristaltic regimes from one another and identify the parameters that control the flow. More importantly, identifying and characterizing the transition from regime 2 to 3 can provide insightful clinical information about the process of transitioning from an effective to an ineffective distention-induced peristalsis. 

Moreover, in this work, we extended \cite{Acharya_2021} 1D model by introducing a relaxation parameter to the flow. The relaxation was implemented distal to the contraction and traveled sinusoidally with the contraction wave. This parameter provides a more realistic representation of flow inside an esophagus. The simulation results with relaxation displayed slightly different shapes than the ones identified by \cite{Acharya_2021} and therefore, the previously used qualitative classification of peristaltic regime type, based on the tube's geometry did not hold. Hence, a qualitative method was needed.

As an approximation, we chose to study a simplified, reduced-order model of our dynamic problem by looking at the solution from the frame of reference of the traveling contraction wave at a specific point in time. In this frame, we derived a non-dimensional friction parameter $\psi_\mu'$ (equation (\ref{eqn: friction_equation_full})) and two scalings for the cross-sectional area of the contraction $A_c''$ and $A_c'''$ (equation \ref{eq:non-dim parameters_Ac})) , which were plotted in two separate figures \ref{fig:fric_vs_Adbpc} and \ref{fig:fric_vs_A3prime_c}.  These plots helped us to extract the following conclusions. First, in peristaltic regimes 1 (Fig. \ref{fig:regime1_demo}) and 2 (Fig. \ref{fig:regime2_demo}), the inertia effects are very small such that the dynamic solution is very close to the static solution. In the regimes 1 and 2 region, the peristaltic contraction is tight and is mostly effective in pushing fluid forward. Because the contraction does not open, an increase in fluid viscosity leads to an increase in fluid resistance. Second, for peristaltic regime 3 (Fig. \ref{fig:regime3_demo}), the relation between fluid viscosity and flow resistance is opposite. The results show that in regime 3, the contraction is open, meaning that any attempt of increasing flow friction by increasing viscosity is countered by an increase in the cross-sectional area of the contraction, which results in decrease of flow resistance. In regime 3, the contraction opens such that fluid flow easily across the contraction.

Lastly, we concluded that the transition process is identified as the region in which the characteristics of both peristaltic regimes 2 and 3 hold, and it can be determined quantitatively based on a critical friction value. Cases classified as peristaltic regime 2, right before the transition, have high fluid resistance parameter but the contraction remains tight. At this stage, fluid has accumulated distal to the contraction and friction forces are high. When the fluid resistance is too high, such that the contraction wave cannot continue moving forward, the contraction cross-sectional area opens, allowing more fluid to go through it, which eventually results in regime 3 configuration. We also observed that although the transition point is determined by a variety of factors, it can be quantified by solely using the peristaltic contraction strength, such that a tighter contraction results in a later transition. 

In the last step of the study, we applied the proposed metric to clinical FLIP data of both control and patient populations which exhibit peristalsis. It was found that all clinical readings fall into the regime 1 and 2 region, which implies that they all have sufficient peristaltic contraction strength. Therefore, we concluded that physiological flows fall in peristaltic regimes 1 and 2. The reason for this is unclear. However, we proposed two possible hypotheses which may explain this observation. The first hypothesis concerns the evolutionally disadvantage of regime 3, since it implies insufficient ability to push food. The second hypothesis concerns fluid viscosity. FLIP data are based on aline solution. Perhaps regime 3 would be observed with solid or semi-solid food. Given the reported values of esophagus stiffness and peristaltic contraction strength, the viscosity of the fluid would need to be about $25,000-50,000$ times the viscosity of water in order to observe regime 3. We also note that the subject population only included individuals who exhibit clear peristalsis and excluded non-peristalsis individuals. Lack of peristalsis is a common trait among disease and disorder groups such as achalasia \citep{Aziz2016,Roman2011,Adler2001}. Therefore, it remains unclear whether non-peristalsis individuals fall within the regime 3 classification. This aspect is left unanswered in this study and should be addressed in future work. If non-peristalsis individuals in fact exhibit regime 3 configuration, the proposed method could be investigated as a potential diagnostic tool which differentiates between healthy and unhealthy swallows.

\bibliographystyle{plainnat}
\bibliography{peristalRegimBib} 

\end{document}